\documentclass[12pt,a4paper]{article}

\usepackage{jheppub}
\usepackage[dvipsnames]{xcolor}
\usepackage{amscd,amsmath,amssymb,amsfonts,xspace,mathrsfs,mathtools}
\usepackage[utf8]{inputenc}
\usepackage{bbm}
\usepackage{graphicx}  
\usepackage{tabu}
\usepackage[color,matrix,arrow]{xy} 
\usepackage{wasysym}
\usepackage{todonotes} 
\usepackage{tensor}   

\newcommand{\be}{\begin{equation}} 
\newcommand{\ee}{\end{equation}} 
\newcommand{\bes}{\begin{equation*}}
\newcommand{\ees}{\end{equation*}}
\newcommand{\sgn}{\mathrm{sgn}}

\newcommand{\CA}{\mathcal{A}}

\newcommand{\cD}{\mathcal{D}}

\newcommand{\CH}{\mathcal{H}}

\newcommand{\CL}{\mathcal{L}} 
\newcommand{\CM}{\mathcal{M}}  
\newcommand{\CN}{\mathcal{N}}

\newcommand{\CR}{\mathcal{R}}

\newcommand{\BR}{\mathbb{R}}

\newcommand{\BP}{\mathbb{P}}

\newcommand{\BQ}{\mathbb{Q}}

\title{Decay channels for double extremal black holes in four dimensions}
\author{Johannes Aspman$^{a}$, Jan Manschot$^{b,c,d}$\\
{\it $^a$Department of Computer Science, Czech Technical University in Prague, Czech Republic}\\
  {\it $^b$School of Mathematics, Trinity College, Dublin 2, Ireland}\\
{\it $^c$Hamilton Mathematical Institute, Trinity College, Dublin 2,
  Ireland}\\
{\it $^d$School of Natural Sciences, Institute for Advanced Study, 1
  Einstein Drive, Princeton, NJ 08540 USA}}

\abstract{  
We explore decay channels for charged black holes with vanishing
temperature in $\mathcal{N}=2$ supersymmetric compactifications of
string theory. If not protected by
supersymmetry, such extremal black holes are expected to decay as a
consequence of the weak gravity conjecture. We concentrate on double
extremal, non-supersymmetric black holes for which the values of the scalar fields are
constant throughout space-time, and explore decay channels for which decay into BPS and anti-BPS
constituents is energetically favorable. We demonstrate the existence
of decay channels at tree level for large families of double extremal
black holes. For specific charges, we also find stable
non-supersymmetric black holes, suggesting recombination of
(anti)-supersymmetric constituents to a non-supersymmetric object.  

}

\preprint{}

\begin{document}
	\maketitle

\section{Introduction}
Black holes with vanishing temperature do not emit Hawking
radiation. Still such extremal black holes are expected to decay, if
not protected by supersymmetry, as a consequence of the Weak Gravity
Conjecture (WGC) \cite{ArkaniHamed:2006dz}. This conjecture
states that gravity is the weakest force in any consistent theory of
quantum gravity. More precisely, the conjecture states that in any
consistent theory of quantum gravity there must exist at least one
object on which the gravitational force is smaller than that due to a
gauge charge \cite{ArkaniHamed:2006dz, Brennan:2017rbf,
  Palti:2019pca}. Consequently, there must be a bound on the mass
of this object in terms of the other charges of the theory. For
example, in four-dimensional gravity with a $U(1)$ gauge field this bound reads \cite{Palti:2019pca}
\begin{equation}
	M \leq M_p\,Q,     
\end{equation} 
where $Q$ is the charge and $M_p$ the
four-dimensional Planck mass. The stronger version of the conjecture,
put forward in \cite{Ooguri:2016pdq}, states that only BPS black holes are
allowed to saturate the weak gravity bound. It is
                                important to note that the WGC is a
                                statement about low-energy effective
                                theories. As such, it does not make
                                qualitative predictions on the exact
                                UV completion of the theory. It is possible that non-supersymmetric states of the UV
                                completion are stable against decay \cite{Long:2021lon}.

Decay channels of black holes have been studied in many cases, see for example
\cite{Kats:2006xp, Cheung:2014vva, Cheung:2018cwt, Aalsma:2019ryi,
  Lee:2018urn, Lee:2018spm, Gendler:2020dfp, Heidenreich:2019zkl, Charles:2019qqt}. A
common approach is to consider higher derivative corrections, and to
demonstrate that their coefficients imply that the ratio $M/Q$
decreases with increasing $Q$. In theories with gauge group $U(1)^b$,
$b>1$, decay of extremal black holes leads to further
conditions, in particular conditions on the convex hull of
charge-to-mass ratios \cite{Cheung:2014vva}.

In the present paper, we will consider decay channels for
four-dimensional black holes in $\CN=2$ supergravity. These theories have
generically multiple $U(1)$ gauge fields, as well as families of
solutions of extremal black holes, both supersymmetric
\cite{Ferrara:1995ih, Ferrara:1996dd}\footnote{See for a comprehensive
  review, for 
  example \cite{Mohaupt:2000mj}.}  as well as non-supersymmetric
\cite{Tripathy:2005qp, Kallosh:2006bt, Ceresole:2007wx}. Both families
involve non-trivial dynamics of the scalar fields, known as the
attractor mechanism, which describes the evolution of the scalar
fields from asymptotic infinity to the horizon. As described above,
the weak gravity conjecture suggests that it is energetically favorable
for the extremal black holes to decay.

We restrict to the simplest class of extremal, supersymmetric black
holes, namely those solutions with constant scalar fields. These
solutions are known as {\it double extremal black
  holes}. In particular, as an indication of possible
  decay we study the ratio of masses between the non-supersymmetric
  double extremal black holes and their constituents.

To avoid non-constant scalar fields for
the constituents, we mostly restrict to decay channels with BPS and anti-BPS
objects.\footnote{Decay channels {\bf 4.2a} and {\bf 4.3e} include
  constituents which are extremal but neither BPS nor anti-BPS.} We are able to demonstrate that this subclass of constituents
provides viable decay channels for large families of extremal black holes. 
In addition, we also explore $R^2$ corrections to these decay channels.

More specifically, we consider compactifications of IIA string theory, with black holes
formed as bound state of D$p$-branes supported on $p$-dimensional cycles of the Calabi-Yau
threefold $X$. Black holes with vanishing D6-brane charge are
amenable to analytic analysis, for example of the attractor
points. We restrict to such black holes in this paper. 
The charge lattice contains supersymmetric cones, which
contain the charges of supersymmetric black holes. The magnetic
charges are carried by D4-branes and are positive for supersymmetric
black holes. These correspond to holomorphic, effective divisors of
$X$. For 
  both supersymmetric \cite{Maldacena:1997de,
    Minasian:1999qn} and non-supersymmetric black holes
  \cite{Dabholkar:2006tb} with positive D4-brane
  charges, the microscopic entropy is rather well understood in terms of the
  Maldacena-Strominger-Witten (MSW) conformal field theory (CFT). Such
a description is not available for generic non-supersymmetric black
holes in IIA supergravity.

By analyzing threshold masses in supergravity,
we explore the stability of non-supersymmetric black holes in two families:
\begin{enumerate}
\item Black holes with positive magnetic charges, but electric
  D0-brane charge opposite to that of supersymmetric black
  holes. The charges of non-supersymmetric states of the MSW CFT 
  lie in this cone of the charge lattice \cite{Kraus:2005vz,
    Kraus:2005zm}. Based on mass ratios, we
    demonstrate in Section \ref{nonBPSdecay}, decay channel {\bf 4.2c},
  that it is energetically favorable for such black holes to decay to a bound state of D0-branes and 
  ``polar'' D0-D4-branes. The latter are themselves
  formed from bound states \cite{Denef:2007vg, deBoer:2008fk}. We
 expect that the quantum-mechanical process for this decay is
 Schwinger pair  creation of D0 and anti-D0-branes in the
 electric-magnetic field of the extremal black holes. Such pair creation in the
 background of Reissner-Nordstr\"om black holes has been discussed by
\cite{Gibbons:1975kk, Hiscock:1990ex}.

Curiously, we
find that including F-term $R^2$ corrections appears
to make these decay channels less favorable. While this could be
viewed as weakening the evidence for the WGC, we expect that a full
analysis of $R^2$ corrections, including D-terms, is likely to be in
better agreement with the predictions of the WGC.

\item Black holes with positive and negative magnetic charges. These
  charges correspond to non-holomorphic divisors. Analogously to five dimensions \cite{Long:2021lon,
    Marrani:2022jpt}, we find that the inequalities for decay depend crucially on geometric data
of the compactification geometry, in particular the triple intersection
numbers\footnote{These numbers are the entries of the
    triple intersection tensor $C_{abc}$ which
    determines the cubic prepotential (\ref{Fpert}) in supergravity.} and the positive cone of divisors. For $b_2(X)=2$ and properly
identifying the full effective cone of the CY threefold, we
establish various valid decay channels at tree level, channels {\bf
  4.4b} in Section \ref{sec:twomodulidecay}.
\end{enumerate}

While we find decay channels for various double extremal 
solutions, we are unable to identify such decay channels for
a few specific
cases of magnetically charged non-supersymmetric states.
We expect that these are stable against decay, and give rise to recombination of
(anti-)supersymmetric constituents to a non-supersymmetric bound
state, as discussed recently also in five dimensions
\cite{Demirtas:2019lfi, Long:2021lon,
  Marrani:2022jpt}. The interpretation in terms of the
  WGC is that these states are required to be purely quantum, or
  microscopic, ie they are elementary particles of the UV completion of the low-energy effective theory.

The paper is organised as follows. In Section \ref{sec:extremality} we
briefly review the background material on $\CN=2$ supergravity and
black holes in four dimensions. Section \ref{sec:attractorSol} is
devoted to finding the attractor solutions for various Calabi-Yau
compactifications. We then discuss various possible decay channels for
these solutions in Section \ref{sec:doubleExtreme}. The connection to
the five-dimensional results is made in Section \ref{sec:4d5d} and we conclude with a brief discussion and outlook in Section \ref{sec:discussion}. In the Appendix we collect some useful formulas and computations.

\section{Extremality and attractors}\label{sec:extremality}
In this Section we review the physical setup we will be working in, namely black hole solutions in four-dimensional $\CN=2$ supergravity. 

\subsection{Review of $\CN=2$ supergravity in four dimensions}
We consider type IIA string theory on a compact Calabi-Yau threefold
(CY3), $X$, or, equivalently, M-theory on a circle times a CY3. 
This gives rise to $\CN=2$ supergravity in four dimensions with
$h^{1,1}(X)$ vector multiplets, with $h^{i,j}(X)$ the Hodge numbers of $X$. 
The bosonic part of the supergravity action takes the form 
\begin{equation}
\begin{split}
  S&=\frac{1}{\kappa_4}\int_{\mathbb{R}^{1,3}}d^4x \sqrt{-G}\left(R-2g_{a\bar b}(\partial t^a)(\partial \bar t^{\bar b})\right.\\
  &\qquad \qquad \left.-f_{AB}(t)F^A_{\mu\nu}F^{B\mu\nu}-\frac{1}{2}\tilde{f}_{AB}(t)F^A_{\mu\nu}F^B_{\rho\sigma}\epsilon^{\mu\nu\rho\sigma}
  \right),
\end{split}
\end{equation} 
where $\kappa_4$ is the four-dimensional Newton's constant, $G$ the
determinant of the space-time metric, and $R$ the Riemann curvature. Moreover, $A,B=0,\dots,h^{1,1}(X)$, and $f_{AB}$,
$\tilde f_{AB}$ are determined in terms of the prepotential $F$ introduced
below \cite{Mohaupt:2000mj, Tripathy:2005qp}. The metric on the
complex moduli space $\CM$ of K\"ahler moduli is
$g_{a\bar b}$.

The complexified K\"ahler moduli $t^a$ are parametrized by the projective coordinates
$X^A$,
\be
\label{Kahlermod}
t^a:=\tfrac{X^a}{X^0}=B^a+iJ^a, 
\ee
where $B^a$ are the B-fields and $J^a$ are the (real)
K{\"a}hler moduli, such that the K{\"a}hler form $J=J^a\omega_a$ with
$\omega_a\in H^{1,1}(X)$ a basis of $H^{1,1}(X)$. The triple intersection numbers of the divisor
dual to $\omega_a$ we denote by $C_{abc}$.  

Let $F(X^A)$ be the prepotential of the theory. This function is
homogenous of degree 2, $F(\lambda X^A)=\lambda^2 F(X^A)$ for $\lambda
\in \mathbb{C}^*$. Using this symmetry, we can consider
the gauge $X^0=1$. 
We will mostly be interested in the large volume limit, $J^a\to
\infty$, where we can neglect higher loop and instanton
corrections. The perturbative prepotential is given by\footnote{Our convention
  for the prepotential follows the literature on extremal black holes
  \cite{Tripathy:2005qp, Kallosh:2006bt}, and differs by a sign from
  some other literature, for example \cite{Mohaupt:2000mj}.}
\begin{equation}
\label{Fpert}
  F(X^A)=\frac{1}{6}C_{abc}\frac{X^aX^bX^c}{X^0}+\frac{1}{24}\frac{1}{64}c_{2,a}\frac{X^a}{X^0}\hat
        A,
 \end{equation}
where $c_{2,a}$ is the second Chern class of $X$, and $\hat A$ is a
chiral background field related to the Weyl multiplet. The latter term
involving $\hat A$ will give rise to a curvature squared (or $R^2$)
correction in the effective action of supergravity. In most of our
discussion, we reduce to tree level and set $c_{2,a}=0$. In the gauge $X^0=1$, we have then
\be 
F(t^a)=\frac{1}{6}C_{abc}t^at^bt^c.
\ee
The K\"ahler potential reads
\begin{equation} 
  \label{KahlerPot}  
	\begin{aligned}
		K(X^A,\bar X^A)&=-\log\left[-i \left((X^A)^*\partial_AF-X^A(\partial_AF)^*\right)\right].
	\end{aligned}
      \end{equation}
      At tree level, this evaluates to 
\begin{equation}
	\begin{aligned}
          K(X^A,\bar X^A)=&-\log\left[i\tfrac{1}{6}C_{abc}(t^a-\bar{t}^a)(t^b-\bar{t}^b)(t^c-\bar{t}^c)\right]=-\log \left[8V_{IIA}\right],
	\end{aligned}
\end{equation} 
with $V_{IIA}$ the tree level CY volume,
\begin{equation}\label{kahlerVolume}
	V_{IIA}=\frac{1}{6}C_{abc}J^aJ^bJ^c.
\end{equation}
This volume is in string units and varies as function of the vector
multiplet moduli \cite{deBoer:2008fk}. For use in
  Sec. \ref{sec:4d5d}, we note that the volume in 11D Planck units belongs
to a hypermultiplet, and is independent of the vector multiplet
moduli. As a result, the volume is fixed in five-dimensional
supergravity. For more details, see also Sec. \ref{sec:4d5d}.

The electric-magnetic charge of a 
$(D0,D2,D4,D6)$ brane bound state is denoted by
\be 
\gamma=(q_0,q_a,p^a,p^0)\in \mathbb{Q}^{2b_2+2}.
\ee
Sometimes it is also useful to consider $\gamma$ as a cohomology
class,
\be
\gamma=p^0 \omega_0+p^a\omega_a+q_a\omega^a+q_0\omega^0\in
\oplus_{j=0}^3\, H^{2j}(X,\mathbb{Q}),
\ee
where $\omega_0$ is the generator of $H^0(X,\mathbb{Z})$, $\omega_a$
is a basis for $H^2(X,\mathbb{Z})$, $\omega^a$ a basis for
$H^4(X,\mathbb{Z})$ and $\omega_0$ the generator for
$H^6(X,\mathbb{Z})$. We will sometimes also consider the
$(\omega_A,\omega^A)$ as basis elements for the Poincar\'e dual
homology.

The superpotential is defined by
\begin{equation}
  \label{KandW}
	\begin{aligned}
		W(\gamma,X^A)&=q_A X^A-p^A\partial_AF,
	\end{aligned}
\end{equation}
while the metric on the complexified K{\"a}hler moduli space, $\CM$, is defined in terms of the K{\"a}hler potential as
\begin{equation}
	g_{a\bar b}\coloneqq\partial_a\partial_{\bar b}K.
\end{equation}

We denote the central charge by $Z(\gamma, X^A,\bar X^A)$, defined as
\be
\label{Ccharge}
Z(\gamma, X^A,\bar X^A)=e^{K/2}\, W(\gamma,X^A).
\ee
Upon a rescaling $X^A\to \lambda X^A$ with $\lambda \in \mathbb{C}^*$, we have $Z\to \lambda/|\lambda|\,Z$.

We note that $W$ is a holomorphic function (\ref{KandW}) of the moduli
$t$ (\ref{Kahlermod}) in the gauge $X^0=1$, and we will also use
$W=W(\gamma,t)$. Similarly, we also use the notation $Z(\gamma,t,\bar
t)=Z(\gamma,t)$ for $Z(\gamma,X^A,\bar
X^A)$ and
elsewhere, omitting the dependence on anti-holomorphic variables where appropriate.

The K{\"a}hler covariant derivative $\nabla_ aZ$ of the central charge reads \cite{Kallosh:2006bt}
\be
\nabla_aZ=\partial_a Z+\tfrac12(\partial_a K)Z.
\ee
We have a simple relation between the covariant derivative of $Z$ and
the derivative of $|Z|$,
\begin{equation}
  \label{|Z|nablaZ}
\begin{aligned}
	\partial_a|Z|=&\bar W^{1/2}\left(\frac{1}{2}\frac{\partial_a W}{W^{1/2}}+\frac{1}{2}(\partial_aK)W^{1/2}\right)e^{K/2}=\frac{1}{2}e^{-i\alpha}\nabla_aZ,	\\
	\bar\partial_{\bar a}|Z|=&\frac{1}{2}e^{i\alpha}\bar\nabla_{\bar a}\bar Z,
\end{aligned}
\end{equation}
where $\alpha$ is the phase of $Z$ \cite{Denef:2000nb} $$e^{i\alpha}=\frac{Z}{|Z|}.$$

Moreover, the covariant derivative acting on $W$ reads
\begin{equation}
  \label{Wcov}
	\nabla_AW=\partial_AW+(\partial_A K)W.
\end{equation}

\subsection{Black hole solutions}
We consider the static spherically symmetric metric \cite{Ferrara:1997tw}
\begin{equation}
  \label{BHmetric}
	ds^2=e^{2U}dt^2-e^{-2U}\left[\frac{c^4}{\sinh^4c\tau}\,d\tau^2+\frac{c^2}{\sinh^2c\tau}d\Omega^2\right],
      \end{equation}
      with $\tau\in (0,\infty)$ a parametrization of the radial
      direction, with $\tau\to 0$ at asymptotic infinity and $\tau\to
      \infty$ near the horizon.

 A one-dimensional Lagrangian describing the radial evolution of $U$,
 $t$ and $\bar t$ as functions of $\tau$ can be derived from the
 two-derivative supergravity action. It is given by
 \cite{Tripathy:2005qp, Kallosh:2006bt}
\begin{equation}
  \label{EffLagr}
	\CL(U,t^a,\bar t^{\bar
          a})=\left(\frac{\partial U}{\partial \tau}\right)^2+g_{a\bar
          a}\frac{\partial t^a}{\partial \tau}\frac{\partial \bar
          t^{\bar a}}{\partial \tau}+e^{2U}V_{BH}(\gamma,t).
\end{equation}
The black hole potential $V_{BH}(\gamma, t)$ is a function of the charges and couplings of the theory and in the case of $\CN=2$ supergravity it takes the form
\begin{equation}\label{VbhDef}
	V_{BH}(\gamma,t)=g^{a\bar b}\,\nabla_a Z\, \bar \nabla_{\bar b} \bar Z+|Z|^2=e^K\left[g^{a\bar b}\nabla_aW(\nabla_b W)^*+|W|^2\right],
\end{equation}
where $Z$ is the central charge (\ref{Ccharge}). 

The Lagrangian is
supplemented by the constraint
\begin{equation}\label{constraint}
	\left(\frac{\partial U}{\partial \tau}\right)^2+g_{a\bar a}\frac{\partial t^a}{\partial \tau}\frac{\partial \bar t^{\bar a}}{\partial \tau}-e^{2U}V_{BH}(\gamma,t)=c^2,
\end{equation}
where $c=2ST$, with $S$ the entropy and $T$ the temperature of the
black hole \cite{Gibbons:1996af}. This condition is a manifestation of
the first law of black hole thermodynamics, stating that the total
energy of the system should be conserved.

The equations of motion from the Lagrangian \eqref{EffLagr} for $U$ and $t$ read
\begin{equation}\label{eoms}
\begin{aligned}
\partial_\tau^2U=& e^{2U}V_{BH}, \\
e^{2U}\frac{\partial V_{BH}}{\partial \bar t^{\bar b}}=&g_{a\bar b}\frac{\partial^2 t^a}{\partial \tau^2}+\left(\frac{\partial g_{a\bar b}}{\partial \bar t^{\bar a}}-\frac{\partial g_{a\bar a}}{\partial \bar t^{\bar b}}\right)\frac{\partial t^a}{\partial \tau} \frac{\partial \bar t^{\bar a}}{\partial \bar\tau}+\frac{\partial g_{a\bar b}}{\partial t^b}\frac{\partial t^a}{\partial \tau}\frac{\partial t^b}{\partial \tau}.
\end{aligned}
\end{equation}
When the moduli space is complex K{\"a}hler we have $\Gamma^a_{b\bar c}=0$ and the second equation simplifies. The Christoffel symbol of the K{\"a}hler metric is
\begin{equation}
\Gamma^a_{bc}=g^{a\bar d}\partial_b g_{c \bar d },
\end{equation}
and we can write
\begin{equation}\label{eomsKahler}
	\frac{\partial^2 t^a}{\partial \tau^2}+\Gamma^a_{bc}\frac{\partial
		t^b}{\partial \tau}\frac{\partial t^c}{\partial
		\tau}=g^{a\bar b}e^{2U}\frac{\partial V_{BH}}{\partial \bar t^{\bar b}}.
\end{equation}
The equations of motion are second order non-linear differential
equations. Thus the initial conditions for $\tau=0$ require the initial values as well as
the initial first derivatives (or velocities).

The Lagrangian can alternatively be written as \cite{Ferrara:1997tw}
\begin{equation}
	\CL=\left(\partial_\tau U\pm
          e^U|Z|\right)^2+\big|\partial_\tau t^a\pm e^{i\alpha}\,e^U g^{a\bar
          b}\bar\nabla_{\bar j}\bar Z\big|^2\mp 2\frac{d}{d\tau}\left(e^U|Z|\right),
\end{equation}
From this it is evident that the first order conditions 
\begin{equation}
  \label{lineareqs}
	\begin{aligned}
		\partial_\tau U&=- e^U|Z|, \\
		\partial_\tau t^a&= - e^{U+i\alpha} g^{a\bar b}\bar\nabla_{\bar b}\bar Z,
	\end{aligned} 
      \end{equation}
      minimize the Lagrangian. Here we fixed the sign, by requiring that
      $e^{-2U}\to \infty$ for $\tau\to \infty$. The constraint (\ref{constraint}) vanishes,
$c=0$, if these linear equations are satisfied. These conditions are
however not necessary for $c=0$.
 
The system described above has a natural interpretation in classical
mechanics as a particle moving in a $(2b_2+1)$-dimensional space, and the constraint (\ref{constraint}) is the
conservation of kinetic plus potential energy. Here the potential
energy is identified with $-e^{2U}V_{BH}$. 
Thus stable extrema correspond to maxima of $V_{BH}$, while minima of $V_{BH}$ are
unstable. Since $\CL$ does not contain a dissipative term, converging
attractor  solutions only occur if the total energy equals the maxima
of $-e^{2U}V_{BH}$ such that the particle approaches the unstable
maximum for $\tau\to 0$. This is in the BPS case ensured by the linear BPS
equations (\ref{lineareqs}), while for the non-BPS second order
equations (\ref{eomsKahler}), such converging
attractor behavior only occurs for the right choice of ``initial
velocity'' $dt^a/d\tau|_{\tau=0}$.

We recall the following terminology:
\begin{itemize}
\item A black hole solution is {\it extremal}, if the constraint (\ref{constraint}) is
  satisfied with $c=0$.
  \item A black hole solution is {\it BPS}, if it satisfies the linear
    equation (\ref{lineareqs}). In supergravity, such solutions are
    supersymmetric. These solutions are a subset of the extremal black
    holes.
    \item An extremal black hole solution is {\it double extremal}, if
      the scalar fields are independent of $r=1/\tau$. Then
      $c^2/\sinh(c\tau)^2\to 1/\tau^2$ in the metric (\ref{BHmetric})
      and $U(\tau)=-\log(1+\sqrt{V_{BH}}\,\tau)$.

      The double
      extremal solutions are a subset of the extremal black
      holes. They come in two types
      \begin{enumerate}
      \item {\it Double extremal BPS black holes}: BPS solutions, satisfying \eqref{lineareqs}, for which
        $\partial t^a/\partial \tau=0$, and thus $\nabla_{a} Z=0$, throughout space-time.
        \item {\it Double extremal non-BPS black holes}: Solutions for
          which $\partial t^a/\partial \tau=0$ throughout
          space-time, but not satisfying \eqref{lineareqs}. As a result,  $\bar \partial_{\bar a} V_{BH}=0$
          but $\nabla_{a} Z\neq 0$.
        \end{enumerate}
  \end{itemize}

At spatial infinity, we have $\tau\to 0$, and $U\to M\tau$, with $M$ the ADM
mass determined at asymptotic infinity. The metric becomes Minkowski
for $r\to \infty$ and the constraint reads 
\begin{equation}\label{attractor}
  \begin{split}
&	M(\gamma,t_\infty, \Sigma)^2-|Z(\gamma,
t_\infty)|^2\\
=&\, c^2+|\nabla_a Z(\gamma,t_\infty)|^2-g_{a\bar a}\Sigma^a\bar \Sigma^{\bar a},
\end{split}
\end{equation}
or
\begin{equation}
\label{massexp}
  M(\gamma,t_\infty,\Sigma)^2=c^2+V_{BH}(\gamma,t_\infty)-g_{a\bar a}\Sigma^a\bar \Sigma^{\bar a},
\end{equation}
where we defined the scalar charge $\Sigma^a\coloneqq
\tfrac{dt^a}{d\tau}\Big|_{\tau=0}$
\cite{Kallosh:2006bt}. 
  
As mentioned above, extremal black holes have zero temperature
and thus $c=0$. The subset of BPS solutions satisfy the linear equations (\ref{lineareqs}), and in particular
\be
\Sigma^a=-g^{a\bar a}\bar\nabla_{\bar a}\bar Z(\gamma, t_\infty).
\ee
So we reproduce the well-known relation between the mass and the
central charge,
\begin{equation}
	M(\gamma, t_\infty)=|Z(\gamma,t_\infty)|. 
\end{equation}

For non-BPS extremal black holes we still have $c=0$, but
$\Sigma^a \neq -g^{a\bar
  a}\bar\nabla_{\bar a}\bar Z(\gamma,t_\infty) $, and therefore also
$M^2\neq |Z|^2$. Eq. (\ref{massexp}) gives an upperbound for the
mass of these black holes, $M^2\leq V_{BH}(\gamma,t_\infty)$. For the case of
double extremal black holes $\Sigma^a=0$, such that we have
$M^2=V_{BH}$. In this paper, we will only be concerned with decay
channels for such double extremal black holes, either BPS or non-BPS.

\subsection{Attractor equations}
The determination of
  the attractor values of the moduli at the horizon,
  $\lim_{\tau\to \infty} t^a(\tau)$, is an important problem since
  these are necessary for the evaluation of the mass and entropy of double
  extremal black holes. Already for the linear BPS equations
(\ref{lineareqs}), this is in general a hard and non-trivial question 
\cite{Shmakova:1996nz} with interesting links to arithmetic geometry 
\cite{Moore:1998pn, Candelas:2019llw, Bonisch:2022mgw}. Recently,
 techniques have also been developed to include non-perturbative genus 0 
instanton contributions \cite{Candelas:2021mwz}. In the non-BPS case,  
Eq. (\ref{eoms}) demonstrates that the values at the horizon $t^a(\infty)$
minimize the effective potential \cite{Goldstein:2005hq, Tripathy:2005qp, Kallosh:2006bt}. 
Therefore, in order to find the attractor solutions we are interested in solving the equations
\begin{equation}\label{minimising}
	\partial_a V_{BH}(\gamma,t_\gamma)=e^{K}\left(g^{b\bar c}(\nabla_a\nabla_bW)\bar\nabla_{\bar c}\bar W+2(\nabla_aW)\bar W\right)=0.
      \end{equation}
      
The BPS attractors minimise the central charge such that $\nabla_A
W=0$, while the non-BPS attractors are the solutions to the above
equations with $\nabla_A W\neq 0$. For the BPS attractor equation,
$\nabla_A W=0$ (\ref{Wcov}), we use that
$$\partial_A K=i\,e^K \left( (X^B)^*\partial_A\partial_BF-(\partial_AF)^*\right).$$   
Then taking the real and imaginary part of $\nabla_A W=0$ gives the
well-known equations
\be
q_A=2{\rm Im}\!\left( e^{K/2}\,\bar Z\,F_A\right),\qquad p^A=2{\rm
  Im}\!\left( e^{K/2}\,\bar Z\,X^A\right). 
\ee

An important quantity when studying the attractor solutions is the matrix
\begin{equation}\label{massmatrix}
	m_{kl}=\frac{1}{2}\partial_k\partial_l V_{BH}(\gamma,t_\gamma),
\end{equation}
and its eigenvalues, which are referred to as the mass matrix and
masses of the scalar fields, respectively, in
\cite{Tripathy:2005qp}. The indices $k,l$ run over the $2h^{1,1}(X)$
real dimensions of the K\"ahler moduli space. 
Here $t_\gamma$ refers to the critical values of
the scalars. An attractor solution is a good solution if the
eigenvalues of $m_{kl}$ are all positive. 

\subsection{D0-D2-D4 black holes}
We will focus on D0-D2-D4 systems, and thus set the D6 brane
charge, $p^0$, to zero. The microscopics of these black holes is described by the MSW CFT
\cite{Maldacena:1997de}, and partition functions can be studied in
detail \cite{deBoer:2006vg, Gaiotto:2006wm, Denef:2007vg,
  Manschot:2010sxc, Alexandrov:2012au}. To establish the attractor equations for both BPS and non-BPS black
holes of this type, we first
specialize various quantities to the case $p^0=0$. In the gauge $X^0=1$, we have for the tree level superpotential \footnote{In other places
in the literature the holomorphic central charge is taken as $-\int_X
e^{-t}\wedge \gamma$. This convention results in a
different sign for $q_0$.}  
\begin{equation} 
	\begin{aligned}
          W=&q_0+q_at^a-\frac{1}{2}C_{abc}p^at^bt^c.
	\end{aligned}
\end{equation}

We introduce various shorthand notations
\begin{equation}
	\begin{aligned}
C_{ab}=&C_{abc}p^c,\qquad C^{ab}C_{bc}=\tensor{\delta}{^a_c},\qquad C_a=C_{abc}p^bp^c,\qquad C=C_{abc}p^ap^bp^c, \\
		L_{ab}=&C_{abc}J^c, \qquad L^{ab}L_{bc}=\tensor{\delta}{^a_c},\qquad  L_a=C_{abc}J^bJ^c,\qquad L=C_{abc}J^aJ^bJ^c,
	\end{aligned}
\end{equation}
as well as the shifted variables
\begin{equation}\label{hatteddefs}
	\hat{q}_0\coloneqq q_0+\frac{1}{2}C^{ab}q_aq_b,\qquad \hat t^a\coloneqq t^a-C^{ab}q_b,
\end{equation}
or, since $C^{ab}q_b$ is real, $\hat B^a=B^a-C^{ab}q_b$. These shifts are motivated by a fractional spectral flow giving an effectively pure D0-D4 system \cite{Gaiotto:2006wm,deBoer:2006vg}. Finally, since $C_{ab}$ induces a quadratic form, of signature $(1,b_2-1)$, for elements $k_1,k_2\in H^4(X,\BR)$, we will make use of the notation $C_{ab}k_1^ak_2^b=k_1\cdot k_2$, and similar.

In Appendix \ref{sec:largevol} we give a few useful explicit formulas for the central charge and the black hole effective potential in terms of the charges and moduli for the D0-D2-D4 system. Using these formulas, we find the BPS condition
\begin{equation}
	\nabla_a W=\frac{i}{4}\frac{L_a}{V_{IIA}}\left[\hat{q}_0-\frac{1}{2}\left((\hat B\cdot \hat B)+2i(J\cdot \hat B)-(J\cdot J)\right)\right]-C_{ab}(\hat B^b+iJ^b)=0.
\end{equation}
This is one set of equations for the real part and one set for the imaginary part. The real part tells us that (since $J^a>0$ in the K{\"a}hler cone)
\begin{equation}
	\frac{L_a}{4V_{IIA}}J\cdot \hat B=C_{ab}\hat B^b\implies \hat{B}^a=0,
\end{equation}
and the imaginary part gives the equation,
\begin{equation}\label{bpscon}
	J^a(\hat q_0+\frac{1}{2}(J\cdot J))=4\,V_{IIA}\,p^a,
\end{equation}
where we used $L^{ab}C_{bc}J^c=p^a$.

Extremising the full potential instead gives the condition that
\begin{equation}
	\begin{split}
		&\partial_aV_{BH}=\frac{i}{4}\frac{L_a}{V_{IIA}}\left[(J\cdot J)^2+(\hat B\cdot \hat B)^2+2(J\cdot \hat B)^2+4\hat{q}_0^2-4\hat q_0(\hat B\cdot \hat B)\right]\\
		&\quad -2iV_{IIA}\left[C_{afg}L^{bf}L^{cg}C_{bd}C_{ce}\hat B^d\hat B^e-2iC_{ab}C_{cd}L^{bc}\hat{B}^d-C_{ab}p^b\right]\\
		&\quad +C_{ab}\left[2(J\cdot \hat B)J^b+2(\hat B\cdot
                  \hat B)\hat B^b-2i(J\cdot J)J^b-2i(J\cdot \hat
                  B)\hat B^b-4\hat q_0\hat B^b\right]. 
	\end{split}
      \end{equation}
      vanishes. The real part now tells us that
\begin{equation}\label{renbpscond}
	2V_{IIA}L^{ab}\hat B^cC_{bc}=J^a(J\cdot\hat B)+\hat B^a(\hat B\cdot\hat B)-2\hat q_0\hat B^a,
\end{equation}
with one solution being $\hat B^a=0$. The imaginary part gives the condition
\begin{equation}\label{imnbpscond}
	\begin{aligned}
		&\frac{L_a}{8V_{IIA}}\left[(J\cdot J)^2+(\hat B\cdot \hat B)^2+2(J\cdot \hat B)^2+4\hat q_0^2-4\hat q_0(\hat B\cdot \hat B)\right]\\
		&=V_{IIA}C_{afg}L^{bf}L^{cg}C_{bd}C_{ce}\hat B^d\hat B^e-V_{IIA}C_{ab}p^b+C_{ab}J^b(J\cdot J)+C_{ab}\hat B^b(J\cdot \hat B)\
	\end{aligned}
\end{equation}
When $\hat B^a=0$ this becomes 
\begin{equation}\label{nbpscon}
	J^a(4\hat{q}_0^2+(J\cdot J)^2)=8V_{IIA}\left((J\cdot J)p^a-V_{IIA}L^{ab}C_{bc}p^c\right).
\end{equation}
In this paper, we focus our attention on this case and refer, in the following, to this equation as the \emph{attractor equation}. For the one-moduli case, we show below that the solutions with $\hat B\neq 0$ are not viable. 

This CFT is chiral and has $(0,4)$
supersymmetry. The entropy (\ref{SBHsusy}) follows from the
Cardy formula with central charges $c_L=C+c_2\cdot p$ and
$c_R=C+c_2\cdot p/2$. For a unitary CFT,
the supersymmetric representations need to satisfy
\begin{equation}
	L_0-\frac{c_L}{24}\geq 0,\qquad \bar L_0-\frac{c_R}{24}\geq 0,
\end{equation}
with $L_0$, $\bar L_0$ being the Virasoro generators. The momentum along the M-theory circle, or the D0 brane charge, is given as the difference between the Virasoro generators,
\begin{equation}
	q_0=L_0-\bar L_0-\frac{c_L-c_R}{24}.
\end{equation}
Putting this together we get a lower bound for $\hat q_0$ for the supersymmetric states,  
\be
\label{q0lb} 
\hat q_0\geq -\frac{c_L}{24}.
\ee
The Cardy formula gives the microscopic entropy for large $|\hat q_0|$ \cite{Maldacena:1997de, Kraus:2005vz, Kraus:2005zm, Dabholkar:2006tb},
\be
\label{SCFT}
\begin{split}
	&\text{BPS:}\qquad \qquad S_{CFT}=2\pi\sqrt{\hat q_0\,c_L/6},\\
	&\text{non-BPS:}\qquad S_{CFT}=2\pi\sqrt{-\hat q_0\,c_R/6}.
\end{split} 
\ee

The equation for the BPS entropy obviously breaks down for $-c_L/24\leq\hat q_0<0$. These states are supersymmetric, but do not
consist of single black hole centers. Instead, they are bound states of multiple constituents. For example,
the states with $\hat q_0=-c_L/24$ are bound states of a D6-brane with an
anti-D6 brane, with the D4-brane charge generated by a flux on the
6-brane worldvolume \cite{Denef:2007vg, deBoer:2008fk}.
Since all states corresponding to $\hat q_0<0$ are bound
states of multiple constituents there is no BPS attractor point
associated to such a total charge. On the other hand, the CFT
states consist of those states at the large volume attractor
point $t^*_\gamma$ \cite{deBoer:2008fk},
\be
(t_\gamma^*)^a=C^{ab}q_b+ ip^a\lambda,
\ee
with $\lambda$ sufficiently large. We refer to the states with $-c_L/24\leq\hat
q_0<0$ as ``polar D0-D4 states'', since these states
  give rise to the so-called polar term in the partition function
  \cite{deBoer:2006vg, Dijkgraaf:2000fq}.

\section{Attractor solutions for one- and two-parameter
  models}\label{sec:attractorSol} In this section we study the attractor solutions for different families of Calabi-Yau manifolds. 

\subsection{The general class of attractor solutions}
There is a general way to solve the attractor equations for any
Calabi-Yau threefold \cite{Shmakova:1996nz, Tripathy:2005qp}. We will start by studying this solution. However, as we will see later this does not give all the non-supersymmetric solutions for the Calabi-Yau manifolds with $h^{1,1}(X)>1$. 
The procedure is to first make the ansatz that $\hat t^a= ip^a z$, for
some real parameter $z$ \cite{Tripathy:2005qp}. For this ansatz, the
attractor equation, \eqref{nbpscon}, reduces to
\begin{equation}\label{miniCond}
	(\hat q_0-\tfrac 16 z^2C)(\hat{q}_0+\tfrac 16 z^2C)=0.
\end{equation}
The first factor corresponds to the BPS solution (satisfying $\nabla_aW=0$) and
the second to the non-BPS one (with $\nabla_a W\neq 0$). We thus have the
two solutions \cite{Shmakova:1996nz}\footnote{Here we choose the sign
  of the solution by requiring that $J^a$ should
  be in the K{\"a}hler cone, and thus positive. For the square root and
  other fractional powers, we will use the convention that the image of a positive real number is
  a positive real number.}  
\begin{equation}\label{attractorSol}
\begin{split} 
  & \text{BPS:}\qquad \qquad \hat t^a_\gamma= ip^a\sqrt{\frac{6\hat q_0}{C}},\\
  & \text{non-BPS:}\qquad \hat t^a_\gamma= ip^a\sqrt{-\frac{6\hat q_0}{C}}.
  \end{split}  
\end{equation}
We will refer to these as the ``general'' solutions in the following, since they hold for any Calabi-Yau.
The K{\"a}hler cone condition tells us that $J^a=\text{Im}\,t^a>0$, which
thus means that we need $p^a>0$ and $\pm\frac{6\hat q_0}{C}>0$ for the
BPS and non-BPS solutions, respectively. The non-BPS solutions
have the same charges as the non-BPS states of the MSW CFT
\cite{Maldacena:1997de}. Therefore modulo decay of multi-center black
holes in the decoupling limit \cite{deBoer:2008fk}, these solutions are captured
by the MSW CFT.

It is possible to determine the effect of $R^2$
  corrections to F-terms for the attractor
values. For the BPS case, a closed expression is available
\cite{LopesCardoso:1998tkj}, following Wald's formalism. 
For the non-BPS attractor values, an order by
order analysis in $c_2$ can be carried out using the entropy formalism
\cite{Sen:2005wa, Sahoo:2006rp, Cardoso:2006xz}. The results for both
cases are
\begin{equation}\label{attractorSolR2}
\begin{split} 
  & \text{BPS:}\qquad \qquad \hat t^a_\gamma= ip^a\sqrt{\frac{6\hat
      q_0}{C+c_2\cdot p}},\qquad \hat A_\gamma=-\frac{64\,e^{-K(t_\gamma,\bar t_\gamma)}}{\overline{Z(\gamma,t_\gamma)}^2},\\
  & \text{non-BPS:}\qquad \hat t^a_\gamma= ip^a\sqrt{-\frac{6\hat
      q_0}{C}}\left(1-\frac{9}{32}\frac{c_2\cdot
      p}{C}+\dots\right),\qquad \hat A_\gamma=-\frac{4\,e^{-K(t_\gamma,\bar t_\gamma)}}{\overline{Z(\gamma,t_\gamma)}^2}.
  \end{split}
\end{equation}
Thus the magnitude of the K\"ahler modulus is reduced in both
cases. It is important to note that we are only
  considering F-term corrections in the above, while
  non-supersymmetric black holes may also be affected by $R^2$
  corrections to D-terms \cite{de2011new}.

\subsection{One-parameter CYs}
Let us now turn to examples of Calabi-Yau threefolds with $h^{1,1}(X)=1$. For ease of notation we define $\kappa\coloneqq C_{111}$, $p\coloneqq p^1$ and $q\coloneqq q_1$. 

For this simple case we can return to the generic equation for the minimising of the potentials. For the BPS case we saw in \eqref{bpscon} that we need $\hat B=0$, the BPS solution for $J$ is then
\begin{equation}
	J^2_\gamma=\frac{6\hat{q}_0}{\kappa p}.
\end{equation}
For the minimising of the full potential we saw that we can have either $\hat B=0$ or $\hat B\neq0$. The first case gives \eqref{nbpscon} and the solutions
\begin{equation}\label{1parametersols}
	J^2_\gamma=\pm\frac{6\hat{q}_0}{\kappa p}.
\end{equation}
This reproduces the generic solution found in \eqref{attractorSol}, with the minus sign again corresponding to the non-BPS solution. If we instead assume $\hat B\neq 0$ we find that \eqref{renbpscond} and \eqref{imnbpscond} give
\begin{equation}
	\begin{aligned}
		\hat B_\gamma^2=&\frac{18\hat q_0}{\kappa p},\\
		J^2_\gamma=&-\frac{24\hat q_0}{\kappa p}.
	\end{aligned}  
\end{equation}
However, this is not a viable attractor solution. One way to see this
is that it gives opposite signs for $\hat B^2$ and $J^2$, even
though both should be strictly positive for a viable solution (i.e. $\hat B$ and $J$ should be real).
It further gives negative eigenvalues for the mass matrix
\eqref{massmatrix}.  
 
We thus conclude that for $h^{1,1}(X)=1$, the only attractor solutions are the ones given by the general solution \eqref{attractorSol}.

\subsection{Two-parameter CICYs with autochthonous divisors}
\label{subsecC1120}
We now turn to considering the case of $h^{1,1}(X)=2$. Since we now
have two classes of divisors we can get new types of behaviour. It is
in general hard to find all solutions to the
full attractor equations (either BPS or non-BPS) for generic two-parameter
CYs. To simplify, we consider only the cases with $\hat B^a=0$, as in (\ref{nbpscon}). Besides the generic solutions \eqref{attractorSol}, we will see that there are further solutions once we lift the assumption that
$\hat{t}^a\propto p^a$. If we look at classes of CYs where certain
intersection numbers vanish, we can more easily solve the general
equations without making this assumption. To this end, let us start
looking at complete intersection Calabi-Yau (CICY) manifolds in $\mathbb{P}^1\times \mathbb{P}^n$. As we discuss
below, these CYs have $C_{111}=C_{112}=0$ and the generators of their
effective cone include an autochtonous divisor.\footnote{Since the generic expressions such as \eqref{nbpscon} are of course symmetric in $J^1$ and
$J^2$ we get the same type of solutions with the indices $1$ and $2$ interchanged.
} This class of solutions correspond to
16 of the CICYs of \cite{Brodie:2021nit}. There are also 10
toric hypersurface Calabi-Yau (THCY) manifolds in
\cite{Brodie:2021nit} that will satisfy the same conditions, and thus have the same solutions for the attractor equations. See also Tables \ref{tab:cicy} and \ref{tab:thcy} of the Appendix.  In this paper we focus the discussion on the CICYs, but the analysis for the THCY gives the same results.

 The CICYs of the class considered here are all constructed in an
 ambient space of the type $\CA=\BP^1\times \BP^n$, for some $n>1$. We
 thus have an embedding of the Calabi-Yau $X$ in $\CA$, $f:X\to
 \CA$. Assuming that the embedding satisfies the conditions of the
 Lefshetz hyperplane theorem, the cohomology of $X$,
 $H^r(X,\mathbb{Q})$ is isomorphic to $H^r(\CA,\mathbb{Q})$ for $r\leq
 2$. Thus in particular, the cohomology of 2-forms is isomorphic, and
 two generators $\omega_j$ are pull-backs from 2-forms on $\CA$,
 $\omega_a=f^* \eta_a$.  Thus, the map $f^*:H^2(\CA,\mathbb{Q})\to H^2(X,\mathbb{Q})$ combined with Poincar\'e
 duality gives a map $\tilde f^*: H_{2n}(\CA,\mathbb{Q})\to H_2(X,\mathbb{Q})$.  
  
 Our main interest is in the effective cone $C(X)\subset
 H_4(X,\mathbb{Z})$.  The effective cone $C(\CA)\subset
 H_{2n}(\CA,\mathbb{Z})$ is spanned by two divisors $\cD_1$ and
 $\cD_2$. We take these to be $\cD_1\simeq \mathbb{P}^n\subset \CA$ and $\cD_2
 \simeq \BP^1\times \CH_{\BP^n}$, with $\CH_{\BP^n}\in H_{2n-2}(\BP^n)$ the hyperplane of
 $\BP^n$. Using $\tilde f^*$ introduced above, we obtain effective
 divisors of $X$ by $D_a=\tilde
 f^*\cD_a$ for $a=1,2$. Clearly, $D_1$ does not self-intersect, and
 therefore the intersection numbers $C_{111}$ and $C_{112}$ of $X$ vanish.

It is {\it not} true in general that $D_1$ and $D_2$ are the
 generators of the effective cone $C(X)$. Generically, the effective
 cone $C(X)$ is enlarged from that generated by the divisors $D_1$ and
 $D_2$ inherited from the ambient space. The effective cone $C(X)$ is instead generated by $D_1$ and an exceptional divisor, $D_3$, of the form $D_3=mD_2-D_1$ for some $m$ \cite{Ottem:2013,Brodie:2021nit}. See Appendix \ref{app:CYdata} for the explicit forms for the different Calabi-Yau manifolds we consider. This third divisor is sometimes referred to as \emph{autochthonous}, since it is not inherited from an effective divisor on the ambient space \cite{Demirtas:2018akl}. Similarly for the THCYs of Table \ref{tab:thcy}, the effective cones are also generated by a third exceptional divisor (together with one of the divisors inherited from the ambient space), of the same form. We will see later that this third divisor plays an important role when studying the possible decays of the non-supersymmetric black holes \cite{Long:2021lon}. 
 
Among the solutions that minimise the effective potential are of
course the general solutions (\ref{attractorSol}). In addition, we
find for this class of CYs the ``particular solution'',
\begin{equation}\label{c111=c112=0Sols}
	\begin{split}
		J^1_{\gamma}&=-\frac{1}{3C_{122}}(3C_{122}p^1+2C_{222}p^2)\sqrt{-\frac{6\hat
                    q_0}{C}},\\
                J^2_{\gamma}&=p^2\sqrt{-\frac{6\hat q_0}{C}}.
	\end{split}
\end{equation}
This solution can be obtained from the general BPS
solution (\ref{attractorSol}) by the map
\begin{equation}
	p^1\mapsto -\frac{1}{3C_{122}}(3C_{122}p^1+2C_{222}p^2),\qquad p^2\mapsto p^2.
\end{equation}
This transformation keeps $|C|$ invariant but changes the sign of $C$,
$C=3C_{122}p^1(p^2)^2+C_{222}(p^2)^3\mapsto -C$. We will discuss in Section
\ref{sec:4d5d} how this solution is related to the 5-dimensional solutions of
\cite{Long:2021lon, Marrani:2022jpt}.

Let us consider the domain of the charges for which
(\ref{c111=c112=0Sols}) can be a proper solution.
For the moduli to be in the K{\"a}hler cone, we need $p^2>0$ and
$-\frac{1}{3C_{122}}(3C_{122}p^1+2C_{222}p^2)>0$. Since $C_{abc}>0$
for all $a,b,c$, except for $C_{111}=C_{112}=0$ and permutations, then $-3C_{122}p^1>2C_{222}p^2>0$, such
that $p^1$ and $p^2$ have opposite signs. As a result, there is no
overlap between the charge domains for (\ref{c111=c112=0Sols}) and the
general solution for which the $p^a$ have the same sign. 
For (\ref{c111=c112=0Sols}), we have furthermore that $C<0$, such that for this solution to be in the
K{\"a}hler cone we also need $\hat q_0>0$ (contrary to the general non-BPS
solution). The three possible attractor solutions for this class of Calabi-Yau manifolds thus live in three separate charge sectors given by
\begin{equation}
	\begin{aligned}
		\text{BPS:}& \quad p^a>0,\, \hat q_0>0,\\
		\text{General non-BPS:}& \quad p^a>0,\, \hat q_0<0,\\
		\text{Particular non-BPS:}& \quad p^2>0,\, -p^1>\frac{2C_{222}}{3C_{122}}p^2>0,\, \hat q_0>0.
	\end{aligned}
\end{equation}

We note here that, if we instead have $C_{122}=C_{222}=0$, we get the same results as above with $C_{222}$ interchanged with $C_{111}$, and $C_{122}$ with $C_{112}$. In particular, even though we focused the discussion around the complete intersection Calabi-Yaus of Table \ref{tab:cicy}, the same results hold for all the toric hypersurfaces of Table \ref{tab:thcy}.

\subsection{Two-parameter families without autochthonous divisors}\label{sec:c111=c222=0sol}
There are two THCY and one CICY in the classification of
\cite{Brodie:2021nit} that do not get an enlarged K\"ahler cone due to the presence of an autochthonous divisor, these all have $C_{111}=C_{222}=0$. See Table \ref{tab:c111c222}. This means that the effective cones of these manifolds are generated simply by $D_1$ and $D_2$, and these generators have no self-intersection. In five dimensions
these allow for stable non-BPS black strings according to the
analysis of \cite{Long:2021lon,Marrani:2022jpt}. As before, we have the
general solutions (\ref{attractorSol}). In addition there are
particular solutions, which are of a different flavor than the class
of CYs in Subsection \ref{subsecC1120}. Finding all solutions of the
attractor equations \eqref{nbpscon} for $J^a$ is complicated. To
obtain solutions, we start by studying the solutions in terms of the charges $p^a$. For the
case of $C_{111}=C_{222}=0$, Eq. \eqref{nbpscon} gives
\begin{equation}\label{c111=c222=0_OtherSol}
	\begin{aligned}
		p^1=&\frac{2\hat q_0 (C_{112}J_1+2C_{122}J_2)(2C_{112}^2J_1^2+C_{112}C_{122}J_1J_2+C_{122}^2J_2^2)}{J_2(C_{112}J_1+C_{122}J_2)\sqrt{H(J_1,J_2)}}, \\
		p^2=&-\frac{2\hat q_0 (2C_{112}J_1+C_{122}J_2)(C_{112}^2J_1^2+C_{112}C_{122}J_1J_2+2C_{122}^2J_2^2)}{J_1(C_{112}J_1+C_{122}J_2)\sqrt{H(J_1,J_2)}},
	\end{aligned}
\end{equation}
where we defined
\begin{equation}
	\begin{aligned}
		H(J_1,J_2)\coloneqq&\, 4C_{112}^6J_1^6+12C_{112}^5C_{122}J_1^5J_2+21C_{112}^4C_{122}^2J_1^4J_2^2\\
		&+22C_{112}^3C_{122}^3J_1^3J_2^3+21C_{112}^2C_{122}^4+12C_{112}C_{122}^5J_1J_2^5+4C_{122}^6J_2^6,
	\end{aligned}
\end{equation}
for brevity. Note that, similar to the solutions we found for $C_{111}=C_{112}=0$, we must have $\sgn(p^1)\neq\sgn(p^2)$ in order for this solution to be in the K{\"a}hler cone. As stated above, the inverted solutions in terms of $J^a$ are generally not tractable, but using the above solutions we can find a few special cases where things simplify. To illustrate this, we set $p^2=-4p^1<0$ and consider the THCY with
$C_{112}=C_{122}=1$, referred to as $(1,1)^{2,29}_{-54}$ in \cite{Brodie:2021nit}. Entry two of Table \ref{tab:c111c222}. For this particular set of charges the solutions for $J^a$ are easy to find,
\begin{equation}\label{c112=c122=1Sol}
	\begin{aligned}
		J^1_{\gamma}=& \left(\frac{239-57\sqrt{17}}{236}\right)^{1/4}\sqrt{\frac{\hat q_0}{p^1}},\\
		J^2_{\gamma}=&\frac{3+\sqrt{17}}{2}\left(\frac{239-57\sqrt{17}}{236}\right)^{1/4}\sqrt{\frac{\hat q_0}{p^1}}.
	\end{aligned}
\end{equation}

\section{Double extremal black holes and decay
  channels}\label{sec:doubleExtreme}

In this section, we consider decay channels for extremal black
holes. An extremal black hole can decay if the sum of the masses of
the decay products $\sum_j M_j$ is smaller or equal to the total
mass $M$ of the black hole under consideration. Thus if the ratio
\be
\label{Rdef}
\CR=\frac{M^2}{(\sum_j M_j)^2},
\ee
is larger than 1 decay is energetically favorable.
The simplest class of extremal black holes for which
this ratio is readily determined are the double extremal black holes, whose mass squared $M^2$ is given by
$V_{BH}$ (\ref{massexp}) and the moduli are constant, given by the attractor values $t_\gamma$.

To avoid non-trivial attractor flows for the decay products, we
consider BPS and anti-BPS objects as constituents. For such objects, 
the mass simplifies and is given by the absolute value of the central
charge, which is also easily determined for non-constant attractor flows.
We thus consider an extremal, non-BPS black hole with charge $\gamma$
as a bound state of BPS and anti-BPS states with charges $\gamma_j$,
such that $\sum_j \gamma_j=\gamma$. For such decay channels, the
ratio $\CR$ (\ref{Rdef}) becomes 
\begin{equation}\label{massratio} 
	\CR(\gamma,\{\gamma_j\})\coloneqq
        \frac{V_{BH}(\gamma,t_\gamma)}{(\sum_j |Z(\gamma_j,t_\gamma)|)^2},
\end{equation}
where $Z(\gamma_j,t_\gamma)$ (\ref{Ccharge}) is the central charge of
the (anti-)BPS state with charge $\gamma_j$ evaluated at the attractor
point $t_\gamma$ for the total charge $\gamma$. If $\CR> 1$, it is
energetically favorable for the
non-BPS state to decay into the (anti)-BPS constituents,
while if $\CR<1$ the state is stable. The case $\CR=1$ could be
considered as a threshold bound state, and we will find various
threshold decay channels.

If all charges correspond to D-branes of the same dimension, the numerator of $\CR$ (\ref{Rdef}) is mathematically the volume squared
of a submanifold in $X$ with homology class $\gamma$, whose volume is at a local
minimum as function of the moduli parametrizing the embedding of the
submanifold. The denominator of $\CR$ (\ref{massratio}) for (anti)-BPS
constituents is the volume squared of the ``piece-wise
calibrated cycle'', that is to say the volume of the linear
combination of holomorphic and anti-holomorphic cycles, whose homology
class adds up to $\gamma$. 

\subsection{The general BPS solutions}
We review and collect a few results for BPS solutions. At the BPS
attractor point $t^a_\gamma$ (\ref{attractorSol}), the exponentiated K\"ahler potential evaluates to
\be
e^{K(t_\gamma,\bar
  t_\gamma)/2}=\frac{1}{\sqrt{8V_{IIA}}}=\frac{1}{2\sqrt{2}} \left(
  \frac{C}{6\hat q_0^3}\right)^{1/4}.
\ee
The holomorphic central charge reads
\be
\label{WBPS}
W(t_\gamma,\gamma)=4\hat q_0.
\ee
For the central charge we thus find
\be
Z(\gamma,t_\gamma)=\left(2\hat q_0C/3\right)^{1/4}.
\ee
And the mass of the
double extremal BPS black hole,
\be
\label{MassBPS}
M_{BPS}=|Z(\gamma,t_\gamma)|=(2\hat q_0C/3)^{1/4}.
\ee
This reproduces the well-known black hole entropy at tree level
$S_{BH}=\pi \sqrt{2\hat q_0C/3}$ (\ref{SCFT}) \cite{Maldacena:1997de}. If we include the $R^2$ correction,
the mass becomes
\be
M_{BPS}=\left(2\hat q_0(C+c_{2}\cdot p)/3\right)^{1/4},
\ee
and the entropy
\be
\label{SBHsusy}
S_{BH}=\pi \sqrt{2\hat q_0(C+c_{2}\cdot p)/3}.
\ee
Again in agreement with the microscopic entropy (\ref{SCFT})
\cite{Maldacena:1997de, LopesCardoso:1998tkj}.

\subsection{The general non-BPS solutions}
\label{nonBPSdecay}
This subsection considers decay channels for non-BPS states with
generic attractor point (\ref{attractorSol}). We will find that these
black holes can decay at tree level to D0's and polar D0-D4 states.

We start by determining various quantitites. Using the general attractor moduli,
$t^a_\gamma=ip^a\sqrt{-\frac{6\hat{q}_0}{C}}+C^{ab}q_b$ (\ref{attractorSol}) of the non-BPS state we can evaluate the K{\"a}hler potential
\begin{equation}
	e^{K(t_\gamma,\bar t_\gamma)/2}=\frac{1}{\sqrt{8V_{IIA}}}=\frac{1}{2\sqrt{2}}\left(-\frac{C}{6\hat{q}_0^3}\right)^{1/4}.
\end{equation}
The central charge function for this moduli for some set of charges
$\tilde\gamma=(\tilde q_0,\tilde q_a,\tilde p^a)$ is then
\begin{equation}\label{Zattr}
	\begin{aligned}
		Z(\tilde\gamma,t_\gamma)=&\frac{1}{2\sqrt{2}}\left(-\frac{C}{6\hat{q}_0^3}\right)^{1/4}\Bigg[\tilde q_0+i\tilde q_ap^a\sqrt{-\frac{6\hat{q}_0}{C}}+\tilde q_a C^{ab}q_b\\
		&-\frac{1}{2}\left(\tilde p^ap^bC_{ab}\frac{6\hat{q}_0}{C}+2i\tilde p^aq_a\sqrt{-\frac{6\hat q_0}{C}}+C_{abc}\tilde p^a C^{bd}C^{ce}q_dq_e\right)\Bigg].
	\end{aligned}
\end{equation}
If we set $\tilde{\gamma}=\gamma$ we get the central charge of the non-BPS black hole at the attractor point,
\begin{equation}\label{Znb}
	Z(\gamma,t_\gamma)=-\frac{1}{\sqrt{2}}(-\tfrac{1}{6}C\hat q_0)^{1/4}.
\end{equation}

We can also calculate the effective potential by using the formulas given in Appendix \ref{appendix} and evaluate them at the attractor point $t^a_\gamma$,
\begin{equation}
	\begin{aligned}
		W(\gamma,t_\gamma)=&-2\hat{q}_0, \\
		\nabla_a W(\gamma,t_\gamma)=&\frac{1}{2}i C_{ab}p^b\sqrt{-\frac{6\hat{q}_0}{C}}, \\
		g_{a\bar b}(t_\gamma,\bar t_\gamma)=&-\frac{1}{8\hat{q}_0C}(2C_{ab}C-3C_{ac}C_{bd}p^cp^d), \\
		g^{a\bar b}(t_\gamma,\bar t_\gamma)=&-\frac{2}{3}\sqrt{-\frac{6\hat q_0^3}{C}}\left(6\sqrt{-\frac{C}{6\hat q_0}}C^{ab}-3p^ap^b\sqrt{-\frac{6}{\hat{q}_0 C}}\right).
	\end{aligned}
      \end{equation}
Note that the absolute value of the central charge $W$ is smaller for
the non-BPS black hole than for the BPS black hole (\ref{WBPS}).
      
One verifies using these equations that the effective potential
evaluates to 
\begin{equation} 
  \label{VBHnonBPS}
	V_{BH}(\gamma,t_\gamma)=M_{non-BPS}^2=16\,\hat q_0^2\, e^{K}=\sqrt{-2\hat{q}_0C/3}.
      \end{equation}
The entropy $S_{BH}=\pi \sqrt{-2\hat{q}_0C/3}$ agrees with the leading
term of the microscopic entropy \eqref{SCFT} \cite{Kraus:2005vz, Kraus:2005zm, Dabholkar:2006tb}.
      
We can compare this to the central charge squared, of the non-BPS object, \eqref{Znb}, which gives 
\begin{equation}
	\frac{V_{BH}(\gamma,t_\gamma)}{|Z(\gamma,t_\gamma)|^2}=4.
\end{equation} 
Thus the tree level mass of this non-BPS black hole is exactly twice the
magnitude of its central charge.
\vspace{.3cm} \\
{\bf Channel 4.2a: Decay into non-BPS constituents}\\
When the $R^2$ corrections from the vector multiplet sector to the
non-BPS mass are included, the mass squared is no longer simply related by
$V_{BH}$ as in (\ref{massexp}). The correction to the non-BPS mass is
determined to first order in \cite{Gruss:2009wm}. It reads 
\be
M_{non-BPS}=\left( -2 \hat q_0 C/3\right)^{1/4}\left(1-\frac{3}{320}\frac{c_2\cdot p}{C}+\dots\right)
\ee
The negative sign does make
such black holes unstable for decay into lighter non-BPS constituents as
suggested by WGC, at least if the charge of the black hole $\gamma$ is
parallel to that of the constituents $\gamma_j$. In that case, the non-BPS black hole
as well as the non-BPS constituents are double extremal. For example, if
we consider the decay $\gamma=n\gamma'\to n\times \gamma'$, we have
for the ratio $\CR$ (\ref{Rdef}), 
\be
\CR(\gamma,\{n\times \gamma'\})=1+(n^2-1)\frac{3}{160}\frac{c_2\cdot p}{C}+\dots> 1.
\ee
Thus for increasing charge the ratio $\CR$ decreases, and non-BPS black
holes of this type can decay to non-BPS states with charge
vectors whose entries are relatively prime. 
\vspace{.3cm} \\
{\bf Channel 4.2b: Decay into D0's and D4's}\label{sec:42b}\\
We proceed by considering decay of the non-BPS bound state into a
number of D0-branes, and separate D4-branes (here we assume that the
D2-brane charge vanishes). Using (\ref{Zattr}), we have for the
central charges 
\begin{equation}
  \label{Z0Za}
	\begin{aligned}
		Z_0\coloneqq Z(\gamma_0,t_\gamma)=&-\frac{1}{2\sqrt{2}}\left(-\frac{Cq_0}{6}\right)^{1/4}, \\
		Z_a\coloneqq Z(\gamma_a,t_\gamma)=&\frac{1}{2\sqrt{2}}\left(-\frac{Cq_0}{6}\right)^{1/4}\left(\frac{3p^aC_a}{C}\right),\qquad (\text{no sum over } a),	
	\end{aligned}
\end{equation}
where the notation is that $Z_0$ is the central charge of a D0-brane
with charge $\gamma_0=(q_0,0,0)$, while $Z_a$ is the central charge function of a D4-brane wrapping the $a$th divisor, i.e. with $\gamma_a=(0,0,\dots,p^a,0,\dots)$. The mass ratio \eqref{massratio} with these constituents is then
\begin{equation}\label{thresholdratio}
\CR(\gamma,\{\gamma_0,\gamma_a\})=\frac{V_{BH}(\gamma,t_\gamma)}{\left(|Z_0|+\sum_{a=1}^{h^{1,1}}|Z_a|\right)^2}=\frac{16}{\left(1+3\sum_{a=1}^{h^{1,1}}\big|\frac{C_ap^a}{C}\big|\right)^2},
\end{equation}
where we again do not make use of the Einstein summation convention in
the denominator. Since $C$ and $C_ap^a$ are both positive for all $a$,
the sum over $a$ evaluates to 1, and we arrive at
\be
\CR(\gamma,\{\gamma_0,\gamma_a\})=1.
\ee
This indicates that the double extremal non-BPS magnetic
black hole could be considered as a threshold bound state of D0 and D4 BPS
constituents.

Similar results were analysed in great detail for the
STU model in \cite{Gimon:2009gk}, where the only non-zero intersection
number is $C_{123}$ and we indeed have that the above ratio is equal
to unity. For the STU model the authors of \cite{Gimon:2009gk} make
use of its U-duality group to argue that the result for the D0-D4
system is generic.

One may wonder whether $R^2$ corrections alter the conclusion of
threshold stability. Including the first order $R^2$ corrections in
(\ref{Z0Za}) using (\ref{attractorSolR2}), one obtains:
\be
	\begin{aligned}
		Z_0=&-\frac{1}{2\sqrt{2}}\left(-\frac{Cq_0}{6}\right)^{1/4}\left(1+\frac{23}{64}\frac{c_2\cdot
                p}{C}+\dots\right), \\
		Z_a=&\frac{1}{2\sqrt{2}}\left(-\frac{Cq_0}{6}\right)^{1/4}\left(\frac{3p^aC_a}{C}\right) \left(1-\frac{23}{192}\frac{c_{2}\cdot
                p}{C}+\dots\right),	
	\end{aligned}
\ee 
with again no sum over $a$. Thus the effect of $R^2$ corrections is
that the mass of the D0-brane increases,
since the $V_{IIA}$ decreases, while the mass of the D4-brane
decreases. Interestingly, if we sum up $|Z_0|+\sum_a
|Z_a|$, the first order corrections cancel exactly. As a result,
while $\CR=1$ (\ref{thresholdratio}) before including corrections, it
becomes 
\be
\label{D0D4R2}
\CR(\gamma,\{\gamma_0,\gamma_p\})=1-\frac{3}{160}\frac{c_2\cdot p}{C}+\dots
\ee
after including corrections. The $R^2$ corrections thus make decay in
this orginally threshold channel less likely. This possibly suggests
that different constituents, possibly non-(anti)-BPS, give viable
decay channels instead of this one.
\vspace{.3cm} \\
{\bf Channel 4.2c: Decay into D0's and D0-D4's}\label{sec:42c}\\
Since the $R^2$ corrections do not improve decay in the channels
studied above, we explore other decay channels at tree level. In particular
decay into a supersymmetric D0-D4 state with charge $\gamma_p=(\tilde q_0,0,p^a)$, and a D0-branes
 with charge $\gamma_0=(q_0-\tilde q_0,0,0)$. Then we have for $Z_0$
 and $Z_p$,
 \be
 \begin{split}
   Z_0&=
   \frac{1}{2\sqrt{2}}\left(-\frac{C}{6q_0^3}\right)^{1/4}(q_0-\tilde q_0), \\
   Z_p&=\frac{1}{2\sqrt{2}}\left(-\frac{C}{6q_0^3}\right)^{1/4}(\tilde
   q_0-3 q_0).
   \end{split}
 \ee
 Now if we determine the ratio $R$ for this decay, we find
 \be
\label{Rqqt0}
 \CR(\gamma,\{\gamma_0,\gamma_p\})=\frac{V_{BH}(\gamma,t_\gamma)}{(|Z_0|+|Z_p|)^2}=\left(1-\frac{\tilde q_0}{2q_0}\right)^{-1}=1+\frac{\tilde q_0}{2q_0}+\dots,
\ee 
where we assumed $|\tilde q_0/q_0|<1$.  Thus we see that if $\tilde q_0$ has the same sign as $q_0$, and thus
 negative, $\CR>1$, and these constituents give rise to a proper decay
 channel. Eq. (\ref{q0lb}) demonstrates that states with $\tilde q_0$
 and $C>0$ indeed exists, these are the ``polar'' D0-D4 states. As explained there, these states are not black hole solutions
 with a single black hole center, but instead bound states of multiple
 constituents, such as D6, anti-D6 and D0-branes. 

 If we include $R^2$ interactions, there will be a competition
 between the negative contribution of (\ref{D0D4R2}) and  the positive
 contribution of (\ref{Rqqt0}), even though the $R^2$ correction
 differs from $\CR=1$ with $O({\rm charge}^{-2})$, and the tree level
 (\ref{Rqqt0}) differs from 1 with $O({\rm charge}^0)$.
Since $q_0$ is unbounded below but $\tilde q_0$ is bounded below by
$-C/24$, the term $\tilde q_0/q_0$ in (\ref{Rqqt0}) can be arbitrarily
small. On the other hand, further $R^2$ corrections are expected
 beyond those of the vector multiplets considered here. So the results here are not conclusive.

We will study the effect of turning on the D2-brane charge for the decay of the one-parameter threefolds below. When we go to threefolds with $h^{1,1}>1$ more possibilities will be available, as we discuss in Sec. \ref{sec:twomodulidecay}.

\subsection{One-parameter models and dyonic black holes}
To reach exact expressions for $\CR$, let us consider the
one-parameter models in this section.  
For these threefolds, there are only the general 
solutions \eqref{1parametersols} to the attractor equations. For the
double extremal case, the mass of the BPS
black hole is given by (\ref{MassBPS}) with $C=\kappa p^3$, while the
mass of the non-BPS black holes is given by \eqref{VBHnonBPS}.

The central charge of the D0-D2-D4 system, for some charges $\tilde\gamma=(\tilde q_0,\tilde q,\tilde p)$, at the non-BPS attractor moduli is, from \eqref{Zattr},
\begin{equation}
	Z(\tilde\gamma,t_\gamma)=e^{K/2}\left[\tilde{q}_0+i(\tilde q p-q\tilde p)A+\frac{q\tilde q}{\kappa p}-\frac{q^2\tilde p}{2\kappa p^2}-3\hat q_0\frac{\tilde p}{p}\right]
\end{equation}
with $A=\sqrt{-\frac{6\hat q_0}{\kappa p^3}}$. 
\vspace{.3cm} \\
{\bf Channel 4.3a: Decay into D0-D4 and anti-D0-D4}\label{sec:43a}\\
If we consider first the case of no D2 branes,\footnote{Equivalently
  we can consider the situation where we have some (anti-)D2
  branes that form a bound state with the D0 branes. We then simply
  put hats on the relevant factors.} since there is only one divisor,
the only possible (anti-)BPS bound state constituents are either that
we have the D0 and D4 branes separate, giving the mass ratio (\ref{Rqqt0}), as we saw in the previous Section.

Alternatively, we can consider decay into a BPS state D0-D4 brane and
anti-BPS state D0-D4-brane, 
\be
\begin{split}
\gamma_1&=(-x q_0,0,z p),\\
\gamma_2&=((1+x)q_0,0,(1-z)p).
\end{split}
\ee
for some $x\geq 0$ and $z\geq1$. This gives the mass ratio
\begin{equation} 
	\CR(\gamma,\{\gamma_1,\gamma_2\})=\frac{V_{BH}(\gamma,t_\gamma)}{(|Z(\gamma_1,t_\gamma)|+|Z(\gamma_2,t_\gamma)|)^2}=\frac{4}{(x+3z-1)^2}\leq 1,
\end{equation}
with the saturation happening at $x=0$, $z=1$, which is the situation
where the D0- and D4-branes are part of separate constituents, as before. 
\vspace{.3cm} \\
{\bf Channel 4.3b: Decay into D0-, D2- and D4-branes}\label{sec:43b}\\
We proceed by letting the D2-brane charges be generic. Then we can, for example,
consider constituents with charges
\be
\begin{split}
&\gamma_0=(q_0,0,0),\\
&\gamma_q=(0,q,0),\\
&\gamma_p=(0,0,p).
\end{split}
\ee
The corresponding central charges are
\begin{equation}
	Z_0\coloneqq  Z(\gamma_0,t_\gamma),\qquad Z_q\coloneqq Z(\gamma_q,t_\gamma)\qquad Z_p\coloneqq Z(\gamma_p,t_\gamma),
\end{equation}
we get the mass ratio
\begin{equation}
  \begin{split}
    \CR(\gamma,\{\gamma_0,\gamma_q,\gamma_p\})&=\frac{V_{BH}(\gamma,t_\gamma)}{(|Z_0|+|Z_q|+|Z_p|)^2}\\
    &=\frac{16(q_0+\frac{q^2}{2kp})^2}{(|q_0|+\big|\frac{q^2}{kp}+iqpA\big|+\big|3
  q_0+2\frac{q^2}{kp}+iqpA\big|)^2}.
\end{split}
\end{equation}
This is equal to one when $q=0$ but smaller than one when $q\neq0$,
such that it does not correspond to an allowed decay channel for the
dyonic black hole.
\vspace{.3cm} \\
{\bf Channel 4.3c: Decay into D0-D2-D4 and anti-D0-D2-D4}\label{sec:43c}\\
Alternatively, we can consider the decay products
with charges 
      \be 
\begin{split}
  \gamma_1&=(-x q_0,y q,z p),\\
\gamma_2&=((1+x) q_0,(1-y) q,(1-z) p),
\end{split}      
      \ee
for some $x\geq0$, $z\geq 1$ and $y\in \BQ$.  The central charges
$Z_j$ are,
\begin{equation}
	Z_1\coloneqq Z(\gamma_1,t_\gamma),\qquad Z_2\coloneqq Z(\gamma_2,t_\gamma).
      \end{equation}
      
This means that $\gamma_1$ is a BPS D0-D2-D4 state while $\gamma_2$ is an anti-BPS state. The mass ratio,
\begin{equation}
	\CR(\gamma,\{\gamma_1,\gamma_2\})=\frac{V_{BH}(\gamma,t_\gamma)}{(|Z_1|+|Z_2|)^2},
\end{equation}
only has a maximum equal to one for certain values of $q$. For example, if $q_0=-50000$, $k=5$ and $p=10$, the maximum is only equal to one if $q\in \{-1,0,1\}$, and this happens when $x=0$, $y=\tfrac{3}{4}$ and $z=1$. If $|q|>1$ the maximum is smaller than one.
\vspace{.3cm} \\
{\bf Channel 4.3d: Decay into D0-D2-D4 and anti-D0-D2-D4}\label{sec:43d}\\
Another possible situation we can consider is where the initial
non-BPS state has no D2 charge but decays into two BPS states with D2
charges, of course adding up to zero. We thus consider two BPS
states with charges $\gamma_1$ and $\gamma_2$ and
$\gamma=\gamma_1+\gamma_2$. The central charges are
\begin{equation}
	Z_1\coloneqq Z(\gamma_1,t_\gamma),\qquad Z_2\coloneqq Z(\gamma_2,t_\gamma),
      \end{equation}
      with
      \be
      \begin{split}
        \gamma_1&=(-x q_0,y q,z p),\\
        \gamma_2&=((1+x)q_0,-y q,(1-z)p),
        \end{split}
      \ee
for $x \geq 0$, $y\in\BQ$ and $z\geq 1$. The mass ratio with these states however also does not reach a value larger than one. It exactly becomes one only when $x=y=0$ and $z=1$, which corresponds to the previous situation where we have no electric charge in the decay constituents and the D0- and D4-branes are split.  
\vspace{.3cm} \\
{\bf Channel 4.3e: Decay into non-BPS D0-D4 and BPS D0-D4}\label{sec:43e}\\
We briefly explore here the case of a decay channel for a D0-D4 double
extremal black hole with charge $\gamma=(q_0,0,p)$, with as constituents an extremal,
but non-(anti)-BPS D0-D4 black hole with charge $\tilde \gamma=(\tilde
q_0,0,\tilde p)$, and a
BPS D0-D4 black hole with charge $\gamma-\tilde \gamma$. The ratio of
interest is then
\be
\CR(\{\tilde \gamma,\gamma-\tilde
\gamma\},t_\gamma)=\frac{V_{BH}(\gamma,t_\gamma)}{(M(\tilde \gamma,t_\gamma,\Sigma)+|Z(\gamma-\tilde \gamma,t_\gamma)|)^2}.
\ee
The upperbound $M(\tilde \gamma,t,\Sigma)^2\leq V_{BH}(\tilde \gamma,t)$ implies,
\be
\CR(\{\tilde \gamma,\gamma-\tilde
\gamma\},t_\gamma) \geq \frac{V_{BH}(\gamma,t_\gamma)}{(\sqrt{V_{BH}(\tilde \gamma,t_{\gamma})} +|Z(\gamma-\tilde \gamma,t_\gamma)|)^2}.
\ee 
Using the variables $x=\tilde q_0/q_0$ and $z=\tilde p/p$, we can
evaluate the rhs exactly, such that
\be\label{ratio_43e}
\CR(\{\tilde \gamma,\gamma-\tilde
\gamma\},t_\gamma)\geq \frac{16}{(2\sqrt{3z^2+x^2}+|3z-x-2|)^2}.
\ee
We find that in case the BPS black hole has positive D0-brane
charge, $q_0-\tilde q_0\geq 0$, the rhs is always $\leq
1$. However, if the D0-D4 state is polar, such that $q_0-\tilde q_0< 0$
the ratio is larger than 1, and thus provides a viable decay channel. This is
similar to what we found for decay channel {\bf 4.2c}. It would be
interesting to study such decay channels from a multi-center
perspective as in \cite{bossard2013non}. We leave a more
in depth analysis of such decay channels for future work.

\subsection{Decay channels for CICYs with autochthonous divisors}\label{sec:twomodulidecay}
Let us now turn to the complete intersection Calabi-Yau models with $h^{1,1}(X)=2$ studied
in Sec. \ref{sec:attractorSol}.\footnote{The story is analogous for the THCY of Table \ref{tab:thcy}.} This family of CY manifolds has
$C_{111}=C_{112}=0$. For this family, we can make contact with the
recent investigations of five-dimensional solutions in
\cite{Long:2021lon,Marrani:2022jpt}, which will be done in Sec. \ref{sec:4d5d}.

We start by considering the ratio of $V_{BH}/|Z|^2$. We find for the
particular attractor solution, \eqref{c111=c112=0Sols}, that $V_{BH}(\gamma,t_\gamma)=16\hat q_0^2$, and we have the ratio
\begin{equation}
	\frac{V_{BH}(\gamma,t_\gamma)}{|Z(\gamma,t_\gamma)|^2}=4.
\end{equation}
This equals the ratio for the general attractor solution \eqref{attractorSol}.
\vspace{.3cm} \\
{\bf Channel 4.4a: Charges spanned by $D_1$ and $D_2$}\label{sec:44a}\\
We proceed by considering decay channels for vanishing electric
charge, $q_a=0$. Thus the total charge reads
$\gamma=q_0+p^1D_1+p^2D_2$ with $D_1$ and $D_2$ divisors. The constraints on the charges for this
attractor point are $p^2>0$, $p^1<-\frac{2C_{222}}{3C_{122}}p^2$ and
$\hat q_0>0$. We can thus set $p^1=-\frac{2C_{222}}{3C_{122}}np^2$ for
some $n>1$. For very large $n$, this charge approaches the cone of
charges populated by BPS black holes.

For the decay channels, we first study
\be
\begin{split}
&  \gamma_0=q_0, \\
&  \gamma_1=p^1\,D_1, \\
&  \gamma_2=p^2\,D_2,
\end{split}
\ee
such that $\gamma=\gamma_0+\gamma_1+\gamma_2$.

\begin{figure}[h!]\center 
\includegraphics[scale=0.8]{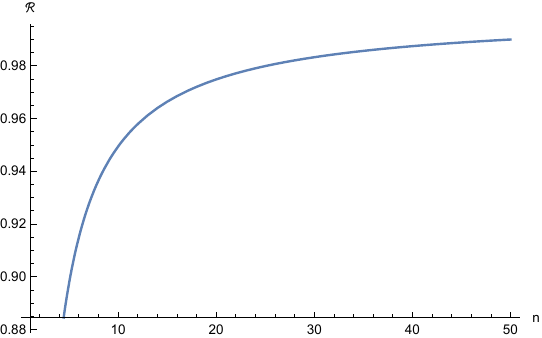}
\caption{Mass ratio of Eq. \eqref{c111=c112=0constituents} as function of the proportionality constant $n$ between the magnetic charges.}
\label{fig:K3optim1}
\end{figure}

In contrast to the general solution we now find
\begin{equation}\label{c111=c112=0constituents}
  \begin{split}
    \CR(\gamma,\{\gamma_0,\gamma_1,\gamma_2\})&=\frac{V_{BH}(\gamma,t_\gamma)}{(|Z(\gamma_0,t_\gamma)|+|Z(\gamma_1,t_\gamma)|+|Z(\gamma_2,t_\gamma)|)^2}\\
    &=\frac{4(1-2n)^2}{(1-4n)^2}<1.
\end{split}
\end{equation} 
This ratio is shown in Fig. \ref{fig:K3optim1}, and we see that it asymptotes to one for large $n$. This is natural to expect, since in the large-$n$ limit the particular solution becomes the BPS one, with a different sign on $p^1$. However, in general we thus see that, we do not have a threshold bound state as for the generic solution. It is therefore interesting to see how this non-BPS state can decay. 
\vspace{.3cm} \\
\noindent {\bf Channels 4.4b: Charges spanned by $D_1, D_2$ and the autochthonous divisor
$D_3$}\label{sec:44b}\\

The charges of the constituents above contained positive linear combinations
of $D_1$ and $D_2$. As mentioned in Section \ref{sec:attractorSol},
the effective cone is generated by an extra, or autochthonous,
effective divisor. This is of the general form $D_3=mD_2-D_1$, for
some $m\geq1$. We have listed these
divisors for various Calabi-Yaus in Appendix \ref{app:CYdata}. Since the cone
extends beyond that generated by $D_1$ and $D_2$, we should allow for
BPS constituents with magnetic charge $\tilde p^1D_1+\tilde p^3 D_3$ with $\tilde p_1,\tilde p_3\geq 0$.

We will study this in what follows. We keep the charge ratio
$p^1=-\tfrac{2C_{222}}{3C_{122}}np^2$ with $n>1$, as before. For the
decay channels, we now consider four constituents with three carrying
a magnetic charge,
\begin{equation}\label{k3constits}
\begin{aligned}
&\gamma_0=x_0q_0,\\
&\gamma_1= x_1q_0+\tilde p^1D_1,\\ 
&\gamma_2=x_2q_0+\tilde p^2D_2,\\ 
&\gamma_3=x_3q_0+\tilde p^3D_3=x_3q_0+\tilde p^3(mD_2-D_1).
\end{aligned}
\end{equation}
From charge conservation we have the following restrictions that must be satisfied
\begin{equation}
	\sum_j x_j=1,\qquad p^1= \,\tilde p^1-\tilde p^3, \qquad
	p^2=\,\tilde p^2+m\tilde p^3.
\end{equation}
We can further assume that $\tilde p^3=z_3p^2$ for some rational number $z_3$. We now find
\begin{equation}
\begin{aligned}
&Z_0=x_0q_0, \\
&Z_1=q_0\left(x_1+3\frac{\frac{C_{122}}{C_{222}}}{2n-1}\left(z_3-\frac{2C_{222}}{3C_{122}}n\right)\right), \\
&Z_2=q_0\left(x_2+\frac{4n-1}{2n-1}(1-mz_3)\right),\\
&Z_3=q_0\left(x_3+\frac{z_3}{2n-1}\left(m(4n-1)-3\frac{C_{122}}{C_{222}}\right)\right).
\end{aligned}
\end{equation}
To have BPS or anti-BPS constituents, it is important that the signs of the coefficients in front of the charges are compatible, in the sense that we must have 
\begin{equation}
\begin{aligned}
\sgn(x_1)=&\sgn(\tilde p^1)=\sgn\left(z_3-\tfrac{2C_{222}}{3C_{122}}n\right), \\
\sgn(x_2)=&\sgn(\tilde p^2)=\sgn(1-mz_3),\\
\sgn(x_3)=&\sgn(z_3).
\end{aligned}
\end{equation}

\begin{figure}[h!]\center
\includegraphics[scale=0.8]{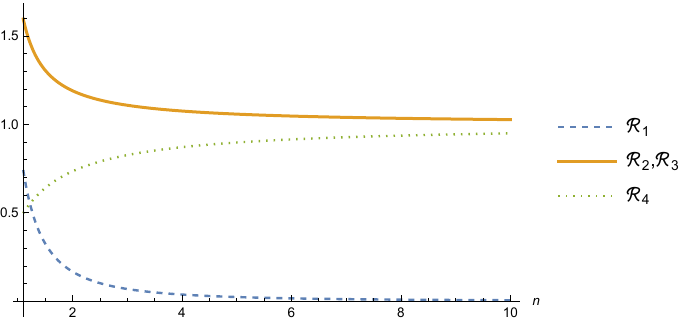}
\caption{Maximised mass ratios for the K3 fibration, or 7887 of Tab. \ref{tab:cicy}, as functions of the proportionality constant $n$. The two ratios $\CR_2$ and $\CR_3$ coincide. }
\label{fig:K3optim2}
\end{figure}

Now, for brevity, let us consider one particular case, namely the K3
fibration \#7887 in Table \ref{tab:cicy}, also studied in
\cite{Long:2021lon}. This has $C_{122}=4$, $C_{222}=2$ and
$D_3=4D_2-D_1$. This means that we now have
$p^1=-\tfrac{n}{3}p^2$. Taking the above analysis and constraints into consideration, and assuming that $x_0\geq0$,\footnote{similar
  results are found when assuming $x_0\leq 0$} we end up with four
different expressions for the mass ratios, viable in four different
regimes for the charges:
\begin{itemize} 
\item For $x_1\geq 0,\,x_2\leq 0\, z_3\geq n/3$:
  $$
\CR_1(\gamma,\{\gamma_j\})=\frac{4(2n-1)^2}{(x_2-2n(1+x_2-8z_3)-4z_3)^2}.
$$
\item For $x_1\leq 0,\, x_2\leq0,\, 1/4\leq z_3\leq n/3$:
  $$
\CR_2(\gamma,\{\gamma_j\})=\frac{4(2n-1)^2}{((2n-1)(x_1+x_2)-2(8n-5)z_3)^2}.
  $$
\item For $x_1\leq0,\, x_2\geq 0, \, 0\leq z_3\leq 1/4$:
$$\CR_3(\gamma,\{\gamma_j\})=\frac{4(2n-1)^2}{(1+2n(x_1-2)-x_1+6z_3)^2}.$$
\item For $x_1\leq 0,\, x_2\geq 0,\, z_3\leq 0$:
  $$
\CR_4(\gamma,\{\gamma_j\})=\frac{4(2n-1)^2}{(x_0+x_2-2n(1+x_0+x_2-8z_3)-4z_3)^2}.
  $$
\end{itemize}
We can maximise these ratios separately over their corresponding domains, and its easy to show that the maximising values for $x_0$, $x_1$, $x_2$ and $z_3$ are given by
\begin{equation}
\begin{aligned}
&\CR_1:\quad x_0=1/4,\, x_1=1/2,\, x_2=0, \, z_3=n/3,\\
&\CR_2: \quad x_0=1/2,\, x_1=0,\,x_2=0,\,z_3=1/4,\\
&\CR_3:\quad x_0=1/4,\, x_1=0,\, x_2=1/2, \, z_3=1/4,\\
&\CR_4: \quad x_0=1/2,\, x_1=0,\,x_2=1/2,\,z_3=0.
\end{aligned}
\end{equation}
The resulting maximised ratios are then functions of the proportionality constant $n$ between $p^1$ and $p^2$, and are shown in Fig. \ref{fig:K3optim2}. We can clearly see that the maximum values for $\CR_2$ and $\CR_3$, which are both given by
\begin{equation}
\CR_2^{\text{max}}=\CR_3^{\text{max}}=\frac{16(1-2n)^2}{(5-8n)^2},
\end{equation}
allows for a ratio larger than unity for any $n$ (although tending to one for large $n$, for the same reason as mentioned above). This means that these regimes allow for the decay of the non-supersymmetric black hole.

\begin{figure}[h!]\center
\includegraphics[scale=0.8]{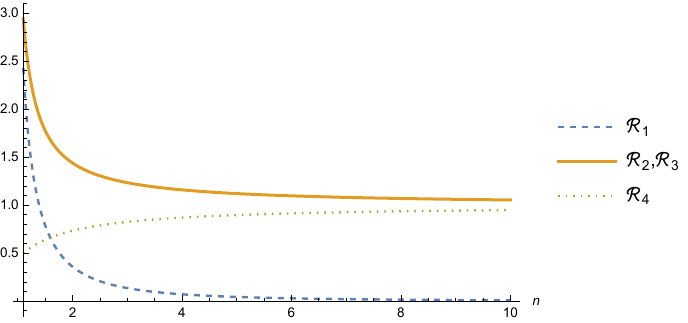}
\caption{Maximised mass ratios for the CICY 7817 of Table \ref{tab:cicy} as functions of the proportionality constant $n$. The two ratios $\CR_2$ and $\CR_3$ coincide completely. }
\label{fig:7817} 
\end{figure}

We can do the same analysis for the other CICYs of Table \ref{tab:cicy}, the result is that 11 out of the 16 gives exactly the same behaviour, i.e. exactly the same maximized ratios as above while the remaining five (7817, 7840, 7858, 7873 and 7885 in Table \ref{tab:cicy}) differ slightly. As an example, the results for the CICY 7817 is shown in Fig. \ref{fig:7817}. Interestingly, we see that for $n$ sufficiently close to 1 we now have two distinct decay channels for the non-supersymmetric black hole. This happens for all the five CICYs not giving the same behaviour as Fig. \ref{fig:K3optim2}.

We can see that the autochthonous divisors are vital for decay at tree
level. If we consider only the cone
generated by positive linear combinations of $D_1$ and $D_2$, we have that the most general BPS-anti-BPS decay products are given by 
\begin{equation}
\begin{aligned} 
&\gamma_1=x q_0-z_1p^1D_1+z_2p^2D_2,\\
&\gamma_2=(1-x)q_0+(1+z_1)p^1D_1+(1-z_2)p^2D_2,
\end{aligned} 
\end{equation}
for some numbers $x,z_2\geq 1$ and $z_1 \geq 0$. The mass ratio \eqref{massratio} is then maximised by the saturating values, $x=z_2=1$, $z_1=0$, giving the result \eqref{c111=c112=0constituents}.
From the graphs we also observe that for large $n$ the ratios approach
1 (or less), which is consistent since for increasing $n$ the charge
of the black hole approaches the cone of charges of  BPS black holes.

The story is different for the general solutions \eqref{attractorSol}. Since the charge of the D0 brane is negative for the general non-BPS solution we think of this as an anti-D0 brane. We consider the same four constituents as before, \eqref{k3constits}. Performing the same analysis as above, dividing the problem into domains depending on the signs of the coefficients and maximising the ratio for each case, we now find that the maximum values are always equal to the threshold value of one.

\subsection{A model without autochthonous divisors}\label{sec:no_aut}
As discussed in Sec. \ref{sec:c111=c222=0sol} there are three two-moduli Calabi-Yau in \cite{Brodie:2021nit} without an autochthonous divisor. These all have $C_{111}=C_{222}=0$. They are the bi-cubic
in $\BP^2\times \BP^2$ with $C_{112}=C_{122}=3$ in \cite{Long:2021lon}
and two THCY with $C_{112}=C_{122}=1$ or $3$ respectively in
\cite{Marrani:2022jpt}. Refs \cite{Long:2021lon, Marrani:2022jpt}
found that non-BPS 5-dimensional black strings are valid and stable
solutions against decay for $p^1/p^2<0$. This gives rise to
recombination of holomorphic and anti-holomorphic cycles mentioned
before. We will discuss in this subsection, non-BPS black holes for
these geometries. 
 
Since there are no autochthonous divisors generating a large cone, the
effective cones of these CYs are simply generated by $D_1$ and $D_2$, the pull
backs of generators of the effective cone of the ambient space. As in
Sec. \ref{sec:attractorSol}, we focus on one example with
$C_{112}=C_{122}=1$ and $p^2=-4p^1$. From the K{\"a}hler condition we
must have that $\hat q_0$ and $p^1$ have the same sign. We thus take
$p^1>0$, which means that $p^2<0$. For the decay channels we consider
decay into a BPS and anti-BPS constituent,
\begin{equation}
\begin{aligned}
\gamma_1=& xq_0+z_1p^1D_1-z_2p^2D_2,\\
\gamma_2=& (1-x)q_0+(1-z_1)p^1D_1+(1+z_2)p^2D_2,
\end{aligned}
\end{equation}
for $x,\,z_1\geq 1$ and $z_2\geq 0$. The mass ratio is then however
always smaller than one, and is maximised by the saturation values
$x=z_1=1$, $z_2=0$, giving $\CR(\gamma,\{\gamma_1,\gamma_2\})\sim
0.83$. Including the possibility of decay into small BPS black holes
with slight negative D0 charge, as in \eqref{Rqqt0}, does not seem to
improve drastically on this bound, and it does not reach above unity
after taking this into consideration. Here we further note that the
maximum values are given when the $C$'s of each constituents are equal
to zero such that for $q_0$ to be smaller than zero as in
\eqref{Rqqt0} we need to consider the $c_{2,a}$ corrections in
$c_L$.

Thus also in this 4-dimensional case, this example suggests that the
spectrum contains a stable non-BPS object. The WGC suggests that these only have small
charges. It would be worthwhile to study more potential decay channels
and to properly include $R^2$ corrections to the mass of the non-BPS solutions.

\section{Lifting to M-theory}\label{sec:4d5d}
The 4-dimensional theory we have considered can be seen as M-theory on
a CY3 $X$ times a circle. From the M-theory perspective the charges of
the D2 and D4 branes then comes from M2 and M5 branes wrapping cycles
in the CY while the D0 brane charge corresponds to momentum along the
circle. If we decompactify the circle we get five-dimensional
theories as studied in \cite{Long:2021lon, Marrani:2022jpt}. Let us
therefore briefly discuss how to relate the solutions. 

In five dimensions the overall volume of $X$ is not dynamical, so we
have one less degree of freedom to work with. This effectively means
that the one-parameter Calabi-Yau manifolds will not give a dynamical
theory in 5d. We therefore restrict to the case of $h^{1,1}(X)=2$. In
five dimensions we will also have that electric objects are pointlike,
i.e. correspond to black holes, while magnetic objects are string
like, i.e, correspond to black strings. So they are treated separately
in \cite{Long:2021lon, Marrani:2022jpt}. We will focus on discussing
the black strings, as these are related to the dyonic solutions we have
studied in this paper. 

Let us denote the five-dimensional vector multiplet (real) scalars by
$\tau^a$ and fix the overall (tree level) volume of the five-dimensional moduli
space,
\begin{equation}
	V_{5d}=\frac{1}{6}C_{abc}\tau^a\tau^b\tau^c,
\end{equation}
 to be equal to one. For the case $C_{111}=C_{112}=0$, it was
found in \cite{Long:2021lon, Marrani:2022jpt} that the
five-dimensional black string attractor solutions are\footnote{Note our conventions
  differ from those of \cite{Marrani:2022jpt}. We have rescaled $C$ by a factor of $-1/6$.}
\begin{equation}\label{5dk3}
	\text{BPS: }\begin{cases}
		\tau^1_{\gamma}=&\frac{p^1}{(C/6)^{1/3}},\\
		\tau^2_{\gamma}=&\frac{p^2}{(C/6)^{1/3}},
	\end{cases}\qquad \text{non-BPS: } \begin{cases}
	\tau^1_{\gamma}=&-\frac{3C_{122}p^1+2C_{222}p^2}{3C_{122}(C/6)^{1/3}},\\
	\tau^2_{\gamma}=&\frac{p^2}{(C/6)^{1/3}}.
\end{cases}
\end{equation}
In \cite{Long:2021lon} the main example of this type is that of the
K3-fibration having $C_{122}=4$ and $C_{222}=2$. 

The relation between the four-dimensional K\"ahler moduli, $J^a$, arising from compactifying type IIA string theory on a CY3, and the five-dimensional vector multiplet scalars, $\tau^a$, coming from compactifying M-theory on the same CY3, is given by
\begin{equation}\label{5D4Drelation}
	\tau^a=\frac{1}{V_{IIA}^{1/3}}J^a,
\end{equation}
where $V_{IIA}$ is, as before, the volume of $X$ in string units \eqref{kahlerVolume} \cite{Gaiotto:2005xt, deBoer:2008fk}. Given the four-dimensional attractor solutions of Sec. \ref{sec:attractorSol}, it is straightforward to evaluate $V_{IIA}$ at the various attractor points,
\begin{equation}
	V_{IIA}=\begin{cases}
		\sqrt{\frac{6q_0^3}{C}},\qquad &\text{BPS \eqref{attractorSol}}, \\
		\sqrt{-\frac{6q_0^3}{C}},\qquad &\text{General non-BPS \eqref{attractorSol}}\\
		\sqrt{-\frac{6q_0^3}{C}},\qquad &\text{Particular non-BPS \eqref{c111=c112=0Sols}}.
	\end{cases}
\end{equation}
It is now straightforward to check that the 4d BPS solutions as well as the particular non-BPS solutions, \eqref{c111=c112=0Sols}, satisfy \eqref{5D4Drelation} when compared to \eqref{5dk3}. Namely, we have 
\begin{equation}\label{4d5dquotient}
	\frac{J^a_{\gamma}}{\tau^a_{\gamma}}=\left|\frac{6q_0^3}{C}\right|^{1/6},
\end{equation}
while for the general non-BPS solution, \eqref{attractorSol},
Eq. \eqref{5D4Drelation} only holds when compared with the 5d BPS
solution. This is expected since the radius of the M-theory circle
goes to infinity and the direction of the momentum does not break
supersymmetry. We thus have  only supersymmetric
attractors at this point. Breaking supersymmetry through reversing of
the orientation of the compactification manifold, such as we have done
for the general solution by flipping the sign of $q_0$, is sometimes
called ``skew-whiffing'' and appears in many places in the literature
\cite{Duff:1986hr, berkooz1999non}. 

In a similar way, we can also study the relation between the solutions for the models with $C_{111}=C_{222}=0$ in four and five dimensions. For the solutions \eqref{c111=c222=0_OtherSol} we start by defining $x=\tfrac{J^1}{J^2}$ and then study the ratio
\begin{equation}
	\frac{p^1}{p^2}=-\frac{x(2C_{112}+C_{122}x)(C_{122}^2+C_{112}C_{122}x+2C_{112}^2x^2)}{(C_{122}+2C_{112}x)(2C_{122}^2+C_{112}C_{122}x+C_{112}^2x^2)}.
\end{equation}
This agrees with the corresponding ratios in the five-dimensional solutions of \cite{Marrani:2022jpt} when setting $C_{112}=C_{122}=1$ or $3$. So these should correspond to the same solutions. For the case of $C_{112}=C_{122}=1$ and $p^2=-4p^1$ we have the solutions \eqref{c112=c122=1Sol} in four dimensions, while the corresponding non-supersymmetric solutions in five dimensions are
\begin{equation}\label{5Dc111=c222=0}
	\begin{aligned}
		\tau^1_{\gamma}=& (-4+\sqrt{17})^{1/3},\\
		\tau^2_{\gamma}=&\left(\frac{7+\sqrt{17}}{2}\right)^{1/3}.
	\end{aligned}
\end{equation}
It is again easy to calculate the 4d K\"ahler volume, $V_{IIA}$, at this attractor point and see that the relation \eqref{5D4Drelation} again hold for these solutions. 

We have thus seen that the 4d particular solutions are the ones that lift to the 5d non-BPS black string solutions of \cite{Long:2021lon, Marrani:2022jpt}.

\section{Discussion}\label{sec:discussion}
Motivated by the weak gravity conjecture, we have studied double extremal attractor black holes in
four-dimensional $\CN=2$ supergravity. In support of the conjecture,
we have demonstrated many decay channels where decay of
non-supersymmetric black holes into BPS and anti-BPS constituents is
energetically favorable. An important aspect of our analysis is the attractor mechanism,
which depends only on the extremality of the black holes and thus allows us
to study both supersymmetric and non-supersymmetric solutions.

Eq. (\ref{Rqqt0}) demonstrates that for the general attractor points
(\ref{attractorSol}) at tree level, non-BPS extremal black holes can decay into D0-branes and ``polar''
D0-D4 branes. We have also explored higher derivative $R^2$
corrections from the vector multiplet sector \cite{Sahoo:2006rp, Cardoso:2006xz, Saraikin:2007jc, Kraus:2005vz,
  Kraus:2005zm}. Curiously, we find with Eq. (\ref{D0D4R2}) that these $R^2$
corrections make these decay channels more stable rather than
unstable. This behavior is untypical for
  $R^2$ corrections, which commonly lead to a larger charge to mass ratio. Notable exceptions are identified for non-supersymmetric
theories in \cite[Section 2.2]{Arkani-Hamed:2021ajd}. Since we studied
in this paper supersymmetric theories, we expect that D-term higher
derivative corrections will further correct the mass formula favoring
decay of non-supersymmetric extremal black holes.
  
Our results are complementary to recent results on black strings in
five-dimensional supergravity \cite{Long:2021lon, Marrani:2022jpt}. Some
qualitative differences between four- and five-dimensional
supergravity is additional electric D0-brane charge, and that the
B-field makes the moduli of
the Calabi-Yau complex. In this
paper we have discussed how these differences affect the results for
black hole decay in four dimensions.

 \begin{table}\begin{center}\footnotesize
 		\begin{tabular}{|c|c|c|c|}
 			\hline
 			Non-BPS solution & Decay & Mass ratio  & Section \\
 			\hline
 			Gen. sol. w/ corr., \eqref{attractorSolR2} & D0s and D4s & $1-\frac{3}{160}\frac{c_2\cdot p}{C}+\dots$ & \ref{sec:42b}b  \\
 			Gen. sol. w/o corr., \eqref{attractorSol} & Polar D0s, and D4s & $1+\frac{\tilde q_0}{2q_0}+\dots$ & \ref{sec:42c}c  \\
 			Gen. sol. w/o corr., \eqref{attractorSol} & Various D0-D2-D4 systems & Always $\leq 1$ & \ref{sec:43a}a-\ref{sec:43d}d \\
 			Gen. sol., \eqref{attractorSol} & BPS D0-D4, and non-BPS D0-D4 & Possibly $>1$& \ref{sec:43e}e \\
 			Part. sol., \eqref{c111=c112=0Sols} & w/o considering autoch. divisor & $\frac{4(1-2n)^2}{(1-4n)^2}<1$ & \ref{sec:44a}a \\
 			Part. sol., \eqref{c111=c112=0Sols} & Including autoch. divisor & $\frac{16(1-2n)^2}{(5-8n)^2}>1$ & \ref{sec:44b}b\\
 			Part. sol., \eqref{c112=c122=1Sol} & No autoch. divisor & $<1$ & \ref{sec:no_aut}\\
 			\hline
 		\end{tabular}
 		\caption{We collect the various results on the decay channels for non-supersymmetric black holes considered in this paper. In Section \ref{sec:42b}b we consider the decay of the general solution, \eqref{attractorSolR2}, into D0- and D4-branes when including $R^2$ corrections, while Secs. \ref{sec:42c}c and \ref{sec:43a}a-\ref{sec:43e}e consider the decay into various D0-D2-D4 and anti-D0-D2-D4 systems without considering $R^2$ corrections. In Sections \ref{sec:44a}a and \ref{sec:44b}b we consider the decay of the particular solutions \eqref{c111=c112=0Sols} into D0-D4 and anti-D0-D4 states whose magnetic charges are spanned either by the divisors inherited directly from the ambient space or by considering the extra autochthonous divisors. For decay to be possible we find that the autochthonous divisor must be considered. Finally, in Sec. \ref{sec:no_aut} we consider Calabi-Yau manifolds that do not have such an autochthounous divisor, we then find that the mass ratio is always smaller than 1 suggesting that these non-BPS states are stable against decay. }
 		\label{tab:results}
 \end{center}\end{table}

We conclude with mentioning a few directions which deserve further
study:
\begin{enumerate}
\item It would be interesting to better understand the stability of black
  holes, whose decay channels to BPS and anti-BPS are only marginally
  unstable at tree level. A better understanding of the $R^2$ corrections to the threshold decay channels {\bf
  4.2b} is desirable. 
\item Studying the decay channels from the perspective of the
2-dimensional CFT could provide important insights in the decay processes.
\item It is desirable to carry out an analogous analysis for extremal
black holes which are not double extremal, that is to say with a
non-trivial flow for the moduli from spatial infinity to the
horizon. To this end, one would need to understand the
non-BPS attractor flows better, possibly including multi-centers \cite{bossard2013non}. This is also of
interest for point 1 above, since one can then explore
decay channels with non-BPS constituents, which may be more
favorable than the ones with BPS and anti-BPS constituents. We have
briefly explored this type of decay in channel {\bf
  4.3e}.

Some possible avenues for progress in this direction is the use of ``fake" supersymmetry
\cite{Ceresole:2007wx}, as well as the results of
\cite{Astefanesei:2007vh} and the solutions of \cite{Tripathy:2017pwi}. 
\end{enumerate}

\acknowledgments
JA would like to thank Cody Long for explaining certain aspects of their paper \cite{Long:2021lon}.
We are also happy to thank Nima Arkani-Hamed, Matthew Rochford and Antoine Vincenti for
discussions. The majority of this work was carried out while JA was a
graduate student in the School of Mathematics, Trinity College Dublin.
During this time, JA was supported by the Government of Ireland Postgraduate
Scholarship Programme GOIPG/2020/910 of the Irish Research Council.
JM is supported by the Laureate Award 15175 “Modularity in Quantum Field
Theory and Gravity” of the Irish Research Council, and the Ambrose Monell Foundation.
 
\appendix
\section{Some useful formulas and notations}\label{appendix}\label{sec:largevol}
Using the notations introduced in Sec. \ref{sec:attractorSol} for the 
D0-D2-D4 system at tree level, with gauge $X^0=1$, we can list various
useful formulas. First, we have for $p^0=0$,
\begin{equation}
	\begin{aligned}
		\partial_aW=&q_a-C_{abc}p^bt^c, \\
		\partial_a K=&\frac{3i}{2}\frac{L_a}{L},
	\end{aligned}
\end{equation}
such that
\begin{equation}\label{nablaW}
	\begin{aligned}
		\nabla_aW=&q_a-C_{abc}p^bt^c+\frac{3iL_a}{2L}W, \\
		(\nabla_aW)^*=&q_a-C_{abc}p^b\bar t^c-\frac{3iL_a}{L}W^*.
	\end{aligned}
\end{equation}
We also need the metric and its inverse
\begin{equation}\label{gij}
	\begin{aligned}
		g_{a\bar b}=& \frac{3}{4L}\left(\frac{3}{L}L_a L_b-2L_{ab}\right),\\
		g^{a\bar b}=& \frac{2L}{3}\left(\frac{3}{L}J^aJ^b-L^{ab}\right),
	\end{aligned}
\end{equation}
where $L^{ab}L_{bc}=\tensor{\delta}{^a_c}$. From this we also find the Christoffel symbols
\begin{equation}
	\Gamma^a_{bc}=\frac{3i}{2L}\left(L_b\delta^a_c+L_c\delta^a_b-L_{bc}J^a\right)-\frac{i}{2}L^{ad}C_{dbc},
\end{equation}
that appear in the equations of motion for the scalar moduli \eqref{eomsKahler}.

As mentioned in Sec. \ref{sec:attractorSol}, at tree level we can express the superpotential as
\begin{equation}
	W(\gamma)= q_0+t^aq_a-\frac{1}{2}C_{abc}p^at^bt^c=\hat q_0-\frac{1}{2}C_{ab}\hat{t}^a\hat{t}^b,
\end{equation}
where $\gamma= (q_0,q_a,p^a,0)$ \cite{Tripathy:2005qp}. Since $t^a=B^a+i J^a$ we can also write this as
\begin{equation}
	W(\gamma)= q_0+(B^a+iJ^a)q_a-\frac{1}{2}C_{ab}(B^aB^b+2iJ^aB^b-J^aJ^b).
\end{equation}
The real and imaginary parts of $W$ are then
\begin{equation}
	\begin{aligned}
		\text{Re}(W)=&q_0+B^aq_a-\frac{1}{2}(B\cdot B-J\cdot J)=\hat q_0+\frac{1}{2}(J\cdot J-\hat B\cdot \hat B), \\
		\text{Im}(W)=&J^aq_a-J\cdot B=-J\cdot \hat B.
	\end{aligned}
\end{equation}
This means that we now have
\begin{equation}\label{Zlv}
	\begin{aligned}
		|W|^2=&\frac{1}{4}(J\cdot J)^2+(J\cdot J)\left(q_0+B^aq_a-\frac{1}{2}B\cdot B\right)+J^aJ^bq_aq_b+(J\cdot B)^2-2J^aq_a(J\cdot B)\\
		&+q_0^2+B^aB^bq_aq_b+\frac{1}{4}(B\cdot B)^2-(B\cdot B)(q_0+B^aq_a)+2q_0B^aq_a\\
		=&\frac{1}{4}(J\cdot J)^2+(J\cdot J)(\hat q_0-\frac{1}{2}(\hat B\cdot \hat B))+(J\cdot \hat B)^2+\hat q_0^2-\hat q_0(\hat B\cdot \hat B)+\frac{1}{4}(\hat B\cdot \hat B)^2.
	\end{aligned}
\end{equation}

The black hole potential can be expressed in a similar way as
\begin{equation}\label{veffGen}
	\begin{aligned}
		e^{-K}V_{BH}=&-4V_{IIA}L^{ab}(q_a-C_{ac}B^c)(q_b-C_{bd}B^d) \\
		&+(J\cdot J)^2-4V_{IIA}(p\cdot J)+2(J\cdot B)^2+2J^aJ^bq_aq_b-4J^aq_a(J\cdot B)\\
		&+4q_0^2+4B^aB^bq_aq_b-4(q_0+B^aq_a)(B\cdot B)+(B\cdot B)^2+8q_0B^aq_a\\
		=&-4V_{IIA}L^{ab}C_{ac}C_{bd}\hat B^c\hat B^d+(J\cdot J)^2-4V_{IIA}(p\cdot J)\\
		&+2(J\cdot \hat B)^2+4\hat q_0^2-4\hat q_0(\hat B\cdot \hat B)+(\hat B\cdot \hat B)^2.
	\end{aligned}
\end{equation}
We can simplify this expression further in special cases. First of all
we consider a D0-D4 system, i.e. setting $q_a=0$. This is equivalent to removing the hats in the above expression, i.e.,
\begin{equation}
	\begin{aligned}
		e^{-K}V_{BH}=&(J\cdot J)^2-4V_{IIA}(p\cdot J)+2(J\cdot B)^2-4V_{IIA}L^{ab}C_{ac}C_{bd}B^cB^d \\
		&+4q^2_0-4q_0(B\cdot B)+(B\cdot B)^2. 
	\end{aligned}
\end{equation}
Alternatively, we can set $\hat B^a=0$, which gives
\begin{equation}
	e^{-K}V_{BH}=(J\cdot J)^2-4V_{IIA}(p\cdot J)+4\hat q_0^2.
\end{equation}

For the one moduli case, $h^{1,1}(X)=1$, it is straightforward to
determine the inverse $L^{ab}$. To state the result for this case, we introduce the shorthand notation $C_{111}=\kappa$, $p^1=p$ and similarly for $J$ and $B$. This gives now
\begin{equation}
	\begin{aligned}
		e^{-K}V_{BH}=&\frac{\kappa^2p^2}{3}\left(J^4+4\hat B^2J^2+3\hat B^4-\frac{12\hat q_0}{\kappa p}\hat B^2+\frac{12\hat q_0^2}{\kappa^2p^2}\right).
	\end{aligned}
\end{equation}

We can also note that $L_{ab}$ is a quadratic form on $H^2(X,\mathbb{R})$ with the signature
$(1,b_2-1)$, similar to $C_{ab}$.

\section{Data of Calabi-Yau manifolds}\label{app:CYdata}
In this Appendix, we collect some data of the Calabi-Yau families of interest to this paper. This data is collected in the Tables \ref{tab:cicy}-\ref{tab:c111c222}. The presentation follows that of \cite{Brodie:2021nit}.

\begin{table}\begin{center}
\begin{tabular}{|c|c|c|c|}
\hline
\# & $(C_{122},C_{222})$ & $n$ & Generators \\
\hline
7806 & (6,6) & 4 & $D_1,\, 2D_2-D_1$ \\
7816 & (8,8) & 5 & $D_1,\,2D_2-D_1$ \\
7817 & (8,12) & 6 & $D_1,\,D_2-D_1$ \\
7819 & (8,16) & 7 & $D_1,\,D_2-D_1$ \\
7822 & (8,8) & 5 & $D_1,\,2D_2-D_1$ \\
7823 & (8,16) & 6 & $D_1,\,D_2-D_1$ \\
7840 & (6,9) & 5 & $D_1,\,D_2-D_1$ \\
7858 & (6,5) & 4 & $D_1,\,2D_2-D_1$ \\
7867 & (6,12) & 6 & $D_1,\,D_2-D_1$ \\
7869 & (6,12) & 5 & $D_1,\,D_2-D_1$ \\
7873 & (6,8) & 5 & $D_1,\,D_2-D_1$ \\
7882 & (6,4) & 4 & $D_1,\,3D_2-D_1$ \\
7885 & (4,5) & 4 & $D_1,\,D_2-D_1$ \\
7886 & (4,8) & 5 & $D_1,\,D_2-D_1$ \\
7887 & (4,2) & 3 & $D_1,\,4D_2-D_1$ \\
7888 & (4,8) & 4 & $D_1,\,D_2-D_1$ \\
\hline
\end{tabular}
\caption{Relevant data for the 16 complete intersection Calabi-Yau
  threefolds with $C_{111}=C_{112}=0$ taken from \cite{Brodie:2021nit}. The number in the first column are the names given in \cite{Brodie:2021nit}. The ambient space is of the form $\CA=\BP^1\times\BP^n$, and the last column lists the generators of the effective cone. }
\label{tab:cicy}
\end{center}\end{table}

\begin{table}\begin{center}
\begin{tabular}{|c|c|c|}
\hline
\# & $(C_{111},C_{112},C_{122},C_{222})$ & Generators \\
\hline
(7,1) & (8,4,0,0)  & $D_2,\, D_1-2D_2$ \\
(8,1) & (8,4,0,0)  & $D_2,\,D_1-D_2$ \\
(8,2) & (8,4,0,0)  & $D_2,\,D_1-D_2$ \\
(9,1) & (0,0,4,2)  & $D_1,\,4D_2-D_1$ \\
(13,1) & (4,5,0,0)  & $D_2,\,D_1-D_2$ \\
(28,1) & (4,2,0,0)  & $D_2,\,D_1-2D_2$ \\
(29,1) & (4,2,0,0)  & $D_2,\,D_1-D_2$ \\
(29,2) & (4,2,0,0)  & $D_2,\,D_1-D_2$ \\
(30,1) & (108,12,0,0)  & ? \\
(32,2) & (108,12,0,0)  & ? \\
\hline
\end{tabular}
\caption{Relevant data for the 10 toric hypersurace Calabi-Yau
  threefolds with either $C_{122}=C_{222}=0$ or $C_{111}=C_{112}=0$
  taken from \cite{Brodie:2021nit}. The numbers in the first column are the names given in \cite{Brodie:2021nit}. The last column lists the generators of the effective cone, where for the last two manifolds the generators of the effective cone are not found in \cite{Brodie:2021nit}.}
\label{tab:thcy}
\end{center}\end{table}

\begin{table}\begin{center}
\begin{tabular}{|c|c|c|}
\hline
\# & $(C_{112},C_{122})$ & Type \\
\hline
7884 & (3,3)  & CICY, $\CA=\BP^2\times \BP^2$ \\
(1,1) & (1,1)  & THCY \\
(5,2) & (3,3)  & THCY \\
\hline
\end{tabular}
\caption{Relevant data for the 3 Calabi-Yau threefolds with $C_{111}=C_{222}=0$ from \cite{Brodie:2021nit}. The numbers in the first column are the names given in \cite{Brodie:2021nit}.}
\label{tab:c111c222}
\end{center}\end{table}

\providecommand{\href}[2]{#2}\begingroup\raggedright\endgroup

\end{document}